\documentclass[12pt]{elsarticle}
\usepackage[utf8]{inputenc}
\usepackage{setspace}
\usepackage[margin=0.75in]{geometry}
\usepackage{graphicx}
\graphicspath{ {./figures/} }
\usepackage[labelfont=bf, font = normalsize]{caption}
\usepackage{subcaption}
\usepackage{amsmath}
\usepackage{mathrsfs}
\usepackage{bm}
\usepackage{xcolor}
\usepackage{array}
\usepackage{appendix}
\usepackage{tikz}
\usepackage{floatrow}
\usepackage{rotating}
\usepackage{multirow,bigdelim}
\usepackage{array, blkarray, makecell}
\usepackage{tabularx}
\DeclareFloatFont{small}{\footnotesize}% "scriptsize" is defined by floatrow, "tiny" not
\newcolumntype{P}[1]{>{\centering\arraybackslash}p{#1}}
\newcommand{\myparagraph}[1]{\paragraph{#1}\mbox{}\\}

\providecommand{\keywords}[1]
{
  \small	
  \textbf{\textit{Keywords:}} #1
}
            
\usepackage{natbib}
\title{%
	\LARGE
	Computational modeling of in-stent restenosis\\
	\large
	Pharmacokinetic and pharmacodynamic evaluation}

\author[1]{Kiran Manjunatha \corref{cor1}}
\ead{manjunatha@ifam.rwth-aachen.de}
\cortext[cor1]{Corresponding author\\kiran.manjunatha@rwth-aachen.de\\ Mies-van-der-Rohe-Str. 1, 52074 Aachen, Germany}
\author[3]{Nicole Schaaps}
\author[2]{Marek Behr}
\author[3]{Felix Vogt}
\author[1]{Stefanie Reese}

\address[1]{Institute of Applied Mechanics, RWTH Aachen University}
\address[2]{Chair for Computational Analysis of Technical Systems, RWTH Aachen University }
\address[3]{Department of Cardiology, Pulmonology, Intensive Care and Vascular Medicine, RWTH Aachen University}
\date{}

\begin{document}

%%%%%% Abstract %%%%%%

\begin{abstract}
Persistence of the pathology of in-stent restenosis even with the advent of drug-eluting stents warrants the development of highly resolved \textit{in silico} models. These computational models assist in gaining insights into the transient biochemical and cellular mechanisms involved and thereby optimize the stent implantation parameters. Within this work, an already established fully-coupled Lagrangian finite element framework for modeling the restenotic growth is enhanced with the incorporation of endothelium-mediated effects and pharmacological influences of rapamycin-based drugs embedded in the polymeric layers of the current generation drug-eluting stents. The continuum mechanical description of growth is further justified in the context of thermodynamic consistency. Qualitative inferences are drawn from the model developed herein regarding the efficacy of the level of drug embedment within the struts as well as the release profiles adopted. The framework is then intended to serve as a tool for clinicians to tune the interventional procedures patient-specifically.\\
\\
\end{abstract}

\maketitle

%%%%%% Main Text %%%%%%

\section{Introduction}

In-stent restenosis (ISR) is one of the undesirable outcomes of percutaneous coronary intervention (PCI). It occurs as a result of the damage sustained by the endothelial monolayer as well as the interior layers of the vessel wall during the balloon angioplasty and subsequent stent implantation procedure. It refers to the progressive re-narrowing of the luminal cross-sectional area leading to ischemic events necessitating revascularization but in extreme cases even resulting in  myocardial infarction. Despite the randomized trials involving drug-eluting stents (DESs) portraying significant reduction in the occurrence of ISR, due to the complexity of the lesions and of patient-specific characteristics, the real-world registries still show a significant rate of ISR occurrence \cite{bucheri2016}. In this regard, the growing need for \textit{in silico} tools to predict ISR based on clinical observations has engendered a wide variety of computational models. 

\subsection{Computational models for ISR}
The earliest efforts to model the mechanism of neointimal hyperplasia, the underlying pathological mechanism in ISR, dates back to the work of \citet{Evans2008}. They employed coupled cellular autonoma (CA) and agent-based modeling (ABM) that interact across multiple temporal and spatial scales to describe the multiphysical process of restenosis, which were further resolved by \citet{tahir2011, Tahir2013ModellingTE, Tahir2014EndothelialRP, Tahir2015AnIS} in terms of capturing the impact of stent design and deployment, functional endothelium, endothelial recovery, and smooth muscle cell migration respectively. \citet{Boyle2010ComputationalSM, Boyle2011InSP} alternatively proposed a hybrid approach wherein a discrete cell-centered-lattice-based approach involving behavioral patterns of cells was assigned to the local environment, while the damage sustained by the arterial wall was quantified using the finite element method (FEM) based on von Mises stresses. \citet{zahedmanesh2012, Zahedmanesh2014AMM} exploited this hybrid approach to unidirectionally couple FEM and ABM to evaluate and optimize stent deployment parameters including strut thickness and strut spacing. Works of \citet{Keshavarzian2018} and \citet{Li2018AFC} presented bidirectionally coupled constructs between FEM and ABM, \citet{Keshavarzian2018} presenting a general framework for arterial growth and remodeling, while \citet{Li2018AFC} specifically focusing on the pathosis of ISR. \citet{Nolan2018AnIO} introduced the damage-induced cell proliferative aspects due to both instantaneous and cyclic loading into the coupled FEM and ABM construct. \citet{Zun2017ACO} incorporated the blood flow dynamics via FEM and coupled them to an ABM for restenotic growth and presented location-specific validation via comparison to micro-CT and histological data from porcine experiments in \citet{Zun2019LocationSpecificCB}. \citet{Corti2021MultiscaleCM} provided a comprehensive overview of pure ABM and the aforementioned hybrid approaches for modeling inflammatory processes in vasculatures.

A separate class of models which rely on a purely continuum-based description of physiological processes involved in ISR has also been proposed. \citet{fereidoo2017} presented a thermodynamically consistent continuum damage and growth model to model ISR, wherein the multiplicative split in the deformation gradient was employed and an evolution equation for the mass growth resulting in neointima was developed. \citet{he2020} then took over the continuum damage and growth model from \citet{fereidoo2017} and enhanced it with a continuum damage description for the plaque present in the artery before stent implantation. They used this setup to evaluate the quantitative influence of arterial overstretch and overlapping stent struts. In both aforementioned works, damage in the arterial wall was considered to be the main driver for restenotic growth. \citet{escuer2019} presented a highly resolved continuum-based model for ISR, tracking the significant mediators through diffusion-reaction equations and coupling them to a continuum growth description developed by \citet{Garikipati2004ACT}. Recently, the authors from our group proposed a fully-coupled Lagrangian FEM framework with coupled advection-reaction-diffusion equations to model the transport of growth mediators linked to a continuum description of growth, incorporating highly resolved chemotactic and haptotactic movements of cellular constituents in \citep{MANJUNATHA2022106166}.

\subsection{Computational models comprising pharmacological influence on ISR}
The advent of DESs necessitated imbibing the influence of pharmacokinetics and pharmacodynamics associated with the drugs embedded in them to the previously developed ISR models. To the best of the knowledge of the authors, only a handful of models have incorporated the pharmacological effects on restenosis. \citet{Caiazzo2011ACA} utilized a lattice Boltzmann flow solver for the blood flow, an ABM for cell dynamics, and a finite difference based drug diffusion scheme to model the multiscale phenomenon of ISR. \citet{Rossi2012BioresorbablePC} on the other hand adopted a continuum description of drug release coupled to the corpuscular approach for neointimal growth presented by \citet{Busini2007MechanisticMO}. In the recent past, \citet{MCQUEEN2022992} have presented a comprehensive modeling framework, built on top of the model presented by \citet{escuer2019}, incorporating highly resolved drug transport and retention mechanisms demarcating specific and non-specific binding processes. They evaluated the effects of stent design, specifically strut configuration and thickness, and the influence of drug mass and release profiles in the ISR outcome. \\
\vspace{0.1in}

\noindent Given the scarcity of models in this context, it is the aim of the authors of this work to extend the modeling framework presented by \citet{MANJUNATHA2022106166} to include endothelium-mediated effects on the process of restenosis in addition to the pharmacological influence of rapamycin-based drugs that are embedded in current generation DESs. The endothelium is modeled as a surface in contrast to the work of \citet{escuer2019}, along the lines of the coupled bulk-surface FEM presented in \cite{Madzvamuse2016TheBF}. Damage-induced mechanisms involved in restenosis, which are the key drivers in the works of \citet{escuer2019} and \citet{MCQUEEN2022992}, are foregone. Rather the disruption in the endothelial integrity brought about by stent implantation is considered to be the driver for restenotic growth. Additionally, the stent is not modeled explicitly as done in \cite{escuer2019} and \cite{MCQUEEN2022992}, but the drug and the pro-inflammatory cytokines are introduced into the system via relevant flux boundary conditions at regions of stent apposition to the arterial wall. Section \ref{section:pathophysiology} gives an overview of the pathological mechanisms that are intended to be modeled. Section \ref{sect_math_model} presents the mathematical model developed, Section \ref{fe_impl} includes a short overview of the finite element implementation, Section \ref{sect_num_eval} considers two numerical examples for the evaluation of the capabilities of the model, and finally, Section \ref{sect_conclusion} provides the conclusion and a brief overview of the open questions that have yet to be handled in terms of modeling ISR.

\section{Pathophysiological overview}\label{section:pathophysiology}

\subsection{Endothelium mediated restenotic growth}
\begin{figure}[htbp]
    \centering
    \includegraphics[scale=0.3]{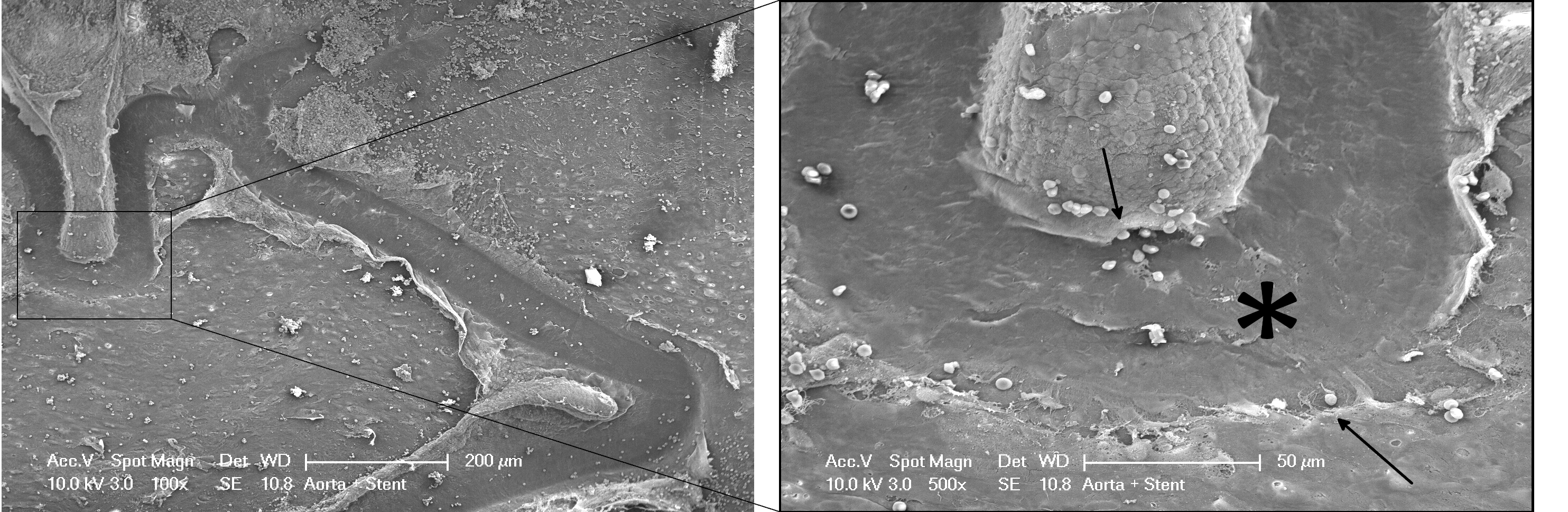}
    \caption{\textbf{microscopy of an explanted rat aorta ( $\bm{\ast}$ - ruptured endothelial region)}}
    \label{fig_EC_microscopy}
\end{figure}

Vascular endothelia are associated with an array of regulatory functions \citep{pober2007}. They modulate local hemostasis by regulating the release of nitric oxide (NO) and prostacyclin. Apart from both compounds being potent vasodilators, they are also associated with inhibitory effects on platelet activation and aggregation. In addition, endothelial cells (ECs) act as a non-permeable barrier against blood-flow contents interacting with the subintimal cellular and extracellular constituents. The process of stent implantation strips the vessel wall off its endothelium thereby exposing the extracellular matrix (ECM) and the smooth muscle cells (SMCs) to the blood flow. Fig. \ref{fig_EC_microscopy} portrays the endothelial rupture observed in a stented section of an explanted rat aorta implanted with a XIENCE-V stent. Due to the scarcity of ECs, the production of NO and prostacyclin is hindered, which leads to a reduction in the anti-platelet activity of the remaining ECs. At this stage, platelets and fibrinogen are rapidly deposited at sites of endothelial denudation kickstarting the inflammatory response of the vessel wall involving cytokines, chemokines, and cellular adhesion molecules (CAMs). The deposited platelets get activated and upon degranulation, release platelet-derived growth factor (PDGF) and transforming growth factor (TGF)-$\beta$ which bring about a proliferative response in the SMCs resulting in neointima formation. Additionally, the upregulated expression of intercellular adhesion molecule (ICAM)-1 and vascular cellular adhesion molecule (VCAM)-1 leads to the recruitment of circulating monocytes at the sites of endothelial injury/dysfunction \citep{clozel1991}. These infiltrating monocytes can adhere to the ECM, the stent surface, and/or the fibrin in the thrombus formed after platelet activation \citep{stewart2009}, and secrete growth factors including PDGF \citep{welt2002, COLOMBO200424}, tumor necrosis factor, and several other mediators from the interleukin family. The enhanced presence of PDGF hence brings about an exacerbated proliferation of SMCs resulting in occlusive restenotic growth. In addition, prolonged exposure of the ECs to low/oscillatory shear from the blood flow shall lead to further accumulation of monocytes and low-density lipoproteins in the subintimal space, which can lead to neoatherosclerosis or exaggeration of neointimal hyperplasia.

\subsection{Pharmacological influence of rapamycin-based drugs}
Rapamycin-based compounds are the most commonly used pharmacological agents that are embedded into modern drug-eluting stents (DESs) and delivered via inorganic polymer layers coated onto the scaffolding stent struts. It is understood that these compounds exert their anti-restenotic effect through two major mechanisms.
\begin{enumerate}
    \item The drugs bind to what are called the FK506 (tacrolimus)-binding proteins, specifically FKBP12. The complex hence formed then binds to the mammalian target of rapamycin (mTOR) protein, a signal transduction kinase, hence arresting the progression of the cell cycle, in the case of both SMCs and ECs, beyond the G1 phase \citep{marx1995}. This results in the maintenance of quiescence of the cellular targets, and therefore in the inhibition of their proliferation. Due to this effect acting indiscriminately on both SMCs and ECs, delayed healing of the endothelium is observed after DES implantation, while also suppression of the proliferation of SMCs happens in congruence. In addition, prolonged rapamycin exposure has been shown to reduce the viability of ECs \citep{barilli2008}, resulting in their apoptosis. Therefore, the usage of rapamycin in DESs engenders effects within the restenotic process which are antithetical. 

    \item The drug presence in the vessel wall after targeted application via DES is observed via \textit{in vitro} cellular assays to be not sufficient enough to bring about the level of reduction in restenotic growth observed angiographically \textit{in vivo}. It is hence hypothesized that the obscure mechanism that results in significant restenosis reduction is the one attributed to the immunosuppressive nature of rapamycin and its analogs. Systemic application of sirolimus has been shown to reduce the level of EC activation, consequently reducing the expression of ICAM-1 and VCAM-1 \citep{DANIEL201779}, which therefore results in a reduced recruitment of monocytes into the subintimal space and their secretion of PDGF. In confluence, these processes result in a significant reduction in restenosis.
\end{enumerate}

\begin{sidewaysfigure}[htbp]
    \centering
    \includegraphics[scale=0.8]{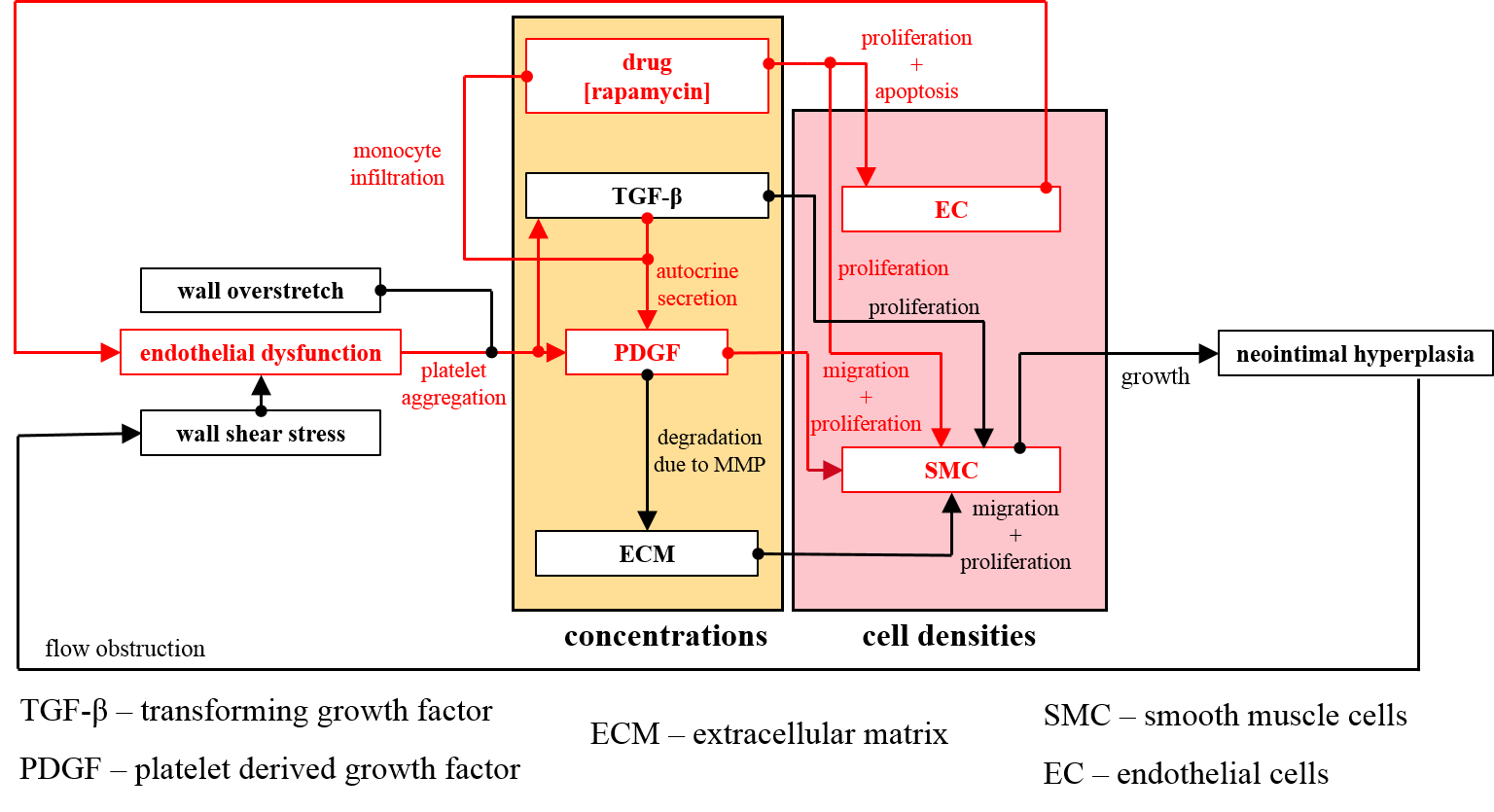}
    \caption{\textbf{Schematic of the restenotic process - updates from \cite{MANJUNATHA2022106166} marked in red}}
    \label{fig_ISR}
\end{sidewaysfigure}

To incorporate the previously described EC modulation of the neointimal growth and the pharmacodynamics of rapamycin, the multiphysics modeling framework presented by \citet{MANJUNATHA2022106166} is extended with two additional field variables, one pertaining to the drug concentration and the other to the EC density. The infiltrating effect of monocytes and their PDGF secretion is indirectly captured by enhancing the source of PDGF through an additional drug-dose-dependent term. Antiproliferative and apoptotic effects of the drug are then coupled back to the platelet aggregation and the inflammatory response. An overview of this updated framework is provided in Fig. \ref{fig_ISR}.

\section{Mathematical modeling}\label{sect_math_model}
To model the pathophysiological and pharmacological processes presented in the previous section, evolution equations are set up for the six mediators of restenosis within the arterial wall and coupled to a continuum growth model. The cellular mediators (SMCs, ECs) of the arterial wall are quantified in terms of cell densities. The extracellular mediators (PDGF, TGF-$\beta$ and ECM) are quantified in terms of their concentrations. The arterial wall, consistent with the media and the adventitia layers, is modeled as an open system allowing for the transfer of cellular and extracellular species into and out of it. Obviously, the blood flow characteristics in the lumen shall affect the restenotic process as described in the pathophysiology section. However, within the current work, only the possibilities of handling them are discussed but not explicitly considered. 

\begin{table}[hbt!]\
\centering
\caption{\textbf{Transport variables}}
\label{transp_vars} 
\begin{tabular}{p{3cm}p{1.5cm}p{4cm}p{2cm}}
\noalign{\hrule height 0.05cm}\noalign{\smallskip}
\\
variable type & variable & associated mediator & units  \\
\\
\noalign{\hrule height 0.05cm}\noalign{\smallskip}
\\
 & $c_{{}_P}$ & PDGF & [mol/mm$^3$]\\
 \\
 concentration & $c_{{}_T}$ & TGF-$\beta$ & [mol/mm$^3$]\\ 
 \\
 & $c_{{}_C}$ & ECM & [mol/mm$^3$]\\
 \\
 & $c_{{}_D}$ & drug & [mol/mm$^3$]\\
 \\
 \hline\noalign{\smallskip}
 \\
cell density & $\rho_{{}_S}$ & SMC & [cells/mm$^3$]\\
\\
& $\rho_{{}_{E}}$ & EC & [cells/mm$^2$]\\
\\
\noalign{\hrule height 0.05cm}\noalign{\smallskip}
\end{tabular}
\end{table}

\subsection{Evolution of the mediators of in-stent restenosis}
The advection-reaction-diffusion equation forms the basis for modeling the transport phenomena governing the evolution of mediators within the arterial wall. The general form for a scalar field $\phi$ is given below:
\begin{equation}\label{ard_eq_general_form}
    \underset{\lower.5em \hbox{\text{rate}}}{\displaystyle{\left.\frac{\partial \phi}{\partial t}\right|_{\bm{x}}}} + \underbrace{\text{\sf div} \left(\phi\,\boldsymbol{v}\right)}_{\lower.3em \hbox{\text{advection}}} = \underbrace{\text{\sf div} \left(k\,\text{\sf grad} \phi\right)}_{\lower.35em \hbox{\text{diffusion}}} + \underbrace{\overset{\text{source}}{\mathcal{R}} - \overset{\text{sink}}{\mathcal{S}}}_{\lower.4em \hbox{\text{reaction}}}.
\end{equation}
\vspace{0.1in}\\
$\bm{v}$ represents the velocity of the medium of transport and $k$  the diffusivity of $\phi$ in the medium. The above general form is valid for arbitrary points within a continuum body in its current configuration represented by the domain $\Omega$. The terms on the right-hand side of Eq. \ref{ard_eq_general_form} must now be particularized for the individual mediators. Table \ref{transp_vars} lists the variables associated with each mediator and their respective units. 

\paragraph{Remark} The cellular and extracellular constituents in the vessel wall undergo constant degradation and regeneration under homeostatic conditions. But within an inflammatory environment (e.g., restenosis, atherosclerosis, fibrosis), this homeostasis is disturbed. For cellular constituents, their programmed death is termed \textit{apoptosis}.   Within the context of the mathematical modeling, it is only intended to model the mechanisms involved in the inflammatory processes thereby ignoring the constant degradative and regenerative processes. 

\subsubsection{Platelet-derived growth factor (PDGF)}
PDGF is a key constituent stored within the $\alpha$-granules of the platelets aggregating at the sites of arterial injury and/or endothelial denudation. 

\begin{figure}[htbp]
    \centering
    \includegraphics[scale=0.6]{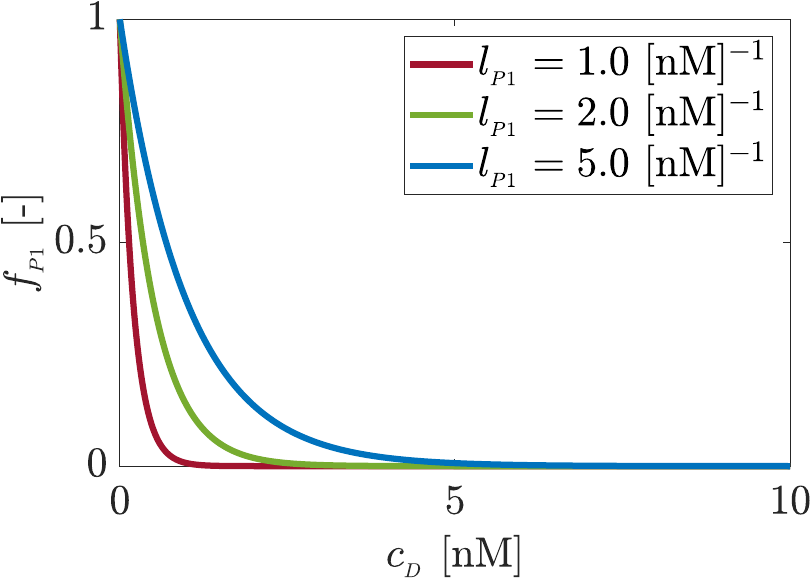}
    \caption{\textbf{Drug-dependent scaling function for the anti-inflammatory effect}}
    \label{fig_fP1}
\end{figure}

\myparagraph{PDGF sources}
In contrast to \citet{MANJUNATHA2022106166}, we herein assume that there are two sources of PDGF. One source ($\mathcal{R}^1_{{}_P}$) is through TGF-$\beta$ ($c_{{}_T}$) bringing about autocrine secretion of PDGF by SMCs ($\rho_{{}_S}$) \citep{gerthoffer2007}. The other (($\mathcal{R}^2_{{}_P}$)), as mentioned in Section \ref{section:pathophysiology}, is due to the monocytes infiltrating the subendothelial space as part of the inflammatory response, their differentiation into macrophages, and the ensuing secretion of PDGF \citep{welt2002, COLOMBO200424, stewart2009, grant2023}. We forego modeling monocyte infiltration with a separate field variable but rather accommodating its effects via TGF-$\beta$-induced autocrine secretion. This is reasonable since TGF-$\beta$ does not contain sources within the arterial wall and its transport can thus be comparable to those of monocytes and differentiated macrophages. A secretion ratio $r_{{}_{\eta}} \in [0,1]$ is introduced in this context, which modulates the relative significance of TGF-$\beta$ induced PDGF autocrine secretion and the effect of monocyte infiltration. Hence the source term for PDGF reads
\begin{equation}\label{PDGFsource_split}
    \mathcal{R}_{{}_P} := (1 - r_{{}_{\eta}})\,\mathcal{R}^1_{{}_P} + r_{{}_{\eta}}\,\mathcal{R}^2_{{}_P}, 
\end{equation}
where, as in \citep{MANJUNATHA2022106166},
\begin{equation}\label{PDGFsource1}
    \mathcal{R}^1_{{}_P} := \eta_{{}_P}\,\rho_{{}_S}\,c_{{}_T},
\end{equation}
where $\eta_{{}_P}$ is termed the autocrine PDGF secretion coefficient.
\myparagraph{Anti-inflammatory pharmacodynamics}
Further, according to \citet{DANIEL201779}, antiproliferative drugs embedded into DESs, particularly rapamycin-based ones (e.g. sirolimus), induce anti-inflammatory effects in the arterial wall. This leads to a significantly lesser infiltration of monocytes into the subendothelial space, thereby inducing a significant anti-restenotic effect in comparison to the direct inhibition of SMC proliferation by the drug itself. To consider this effect, the latter source term ($\mathcal{R}^2_{{}_P}$) is complemented with a factor $f_{{}_{P1}}$, which is a function of the drug concentration ($c_{{}_D}$), which accounts for the reduction in inflammation due to the drug presence, i.e., 
\begin{equation}\label{PDGFsource2}
    \mathcal{R}^2_{{}_P} := f_{{}_{P1}}(c_{{}_D})\, \eta_{{}_P}\,\rho_{{}_S}\,c_{{}_T}.
\end{equation}
\noindent Utilizing Eqs. \ref{PDGFsource1} and \ref{PDGFsource2} in Eq. \ref{PDGFsource_split}, the source term for PDGF hence reads 
\begin{equation}\label{pdgfsource}
\mathcal{R}_{{}_{P}} :=  \left((1 - r_{{}_{\eta}})\, + r_{{}_{\eta}}\,f_{{}_{P1}}(c_{{}_D}) \right) \,\eta_{{}_P}\,\rho_{{}_S}\,c_{{}_T},
\end{equation}
wherein $f_{{}_{P1}}$ is chosen to be of the form
\begin{equation}
    f_{{}_{P1}}(c_{{}_D}) := \displaystyle{\exp ({-l_{{}_{P1}}\,c_{{}_D}})} \in [0,1],
\end{equation}
$l_{{}_{P1}}$ being a parameter that controls the dose-dependent anti-inflammatory effect of the drug (see Fig. \ref{fig_fP1}). This choice is motivated by the fact that rapamycin-based drugs rapidly decrease the expression of cellular adhesion molecules (CAMs) that attract the monocytes into the vessel wall.

\vspace{0.1in}
\noindent The sink term for PDGF is carried over from \citep{MANJUNATHA2022106166}, which is reiterated below for completeness and is given by
\begin{equation}\label{pdgfsink}
    \mathcal{S}_{{}_{P}} := \varepsilon_{{}_P} \,f_{{}_{P2}}(c_{{}_T})\,\rho_{{}_{S}}\, c_{{}_{P}},
\end{equation}
where $\varepsilon_{{}_P}$ is the PDGF receptor internalization coefficient.  The sigmoidal function (see \citep{MANJUNATHA2022106166})
    \begin{equation}
        f_{{}_{P2}}(c_{{}_T}) := \frac{1}{1 + \exp\left({l_{{}_{P2}}\left(c_{{}_T} - c_{{}_{T,th}}\right)}\right)}\,\,\,\in [0,1]
        \label{tgf_scal}
    \end{equation}
captures the scarcity in PDGF receptors induced by TGF-$\beta$ beyond a certain threshold $c_{{}_{T,th}}$, and $l_{{}_{P2}}$ controls steepness of this drop in PDGF receptors.
\vspace{0.1in}\\
\noindent PDGF is considered freely diffusive in the arterial wall, which is modeled by a standard diffusion term as in Eq. \ref{ard_eq_general_form}. The particularized advection-reaction-diffusion equation for PDGF, by utilizing Eqs. \ref{ard_eq_general_form}, \ref{pdgfsource} and \ref{pdgfsink}, hence reads
\begin{equation}\label{pdgf_bal}
    \displaystyle{\left.\frac{\partial c_{{}_P}}{\partial t}\right|_{\bm{x}}} + \text{\sf div} \left(c_{{}_P}\,\boldsymbol{v}\right) = \underbrace{\text{\sf div} \left(D_{{}_{P}}\,\text{\sf grad}\,c_{{}_{P}}\right)}_{\text{diffusion}} + \underbrace{\left((1 - r_{{}_{\eta}}) + r_{{}_{\eta}}\, f_{{}_{P1}}(c_{{}_D})\right)\eta_{{}_P} \rho_{{}_{S}}\, c_{{}_{T}}}_{\substack{\text{secretion by SMCs}\\\text{and macrophages}}} - \underbrace{\varepsilon_{{}_P} \,f_{{}_{P2}}(c_{{}_T})\,\rho_{{}_{S}}\, c_{{}_{P}}}_{\substack{\text{receptor}\\\text{internalization}}},
\end{equation}
where $D_{{}_P}$ refers to the diffusivity of PDGF in the arterial wall.

\begin{table}[htbp!]\
\centering
\caption{\textbf{Model parameters - PDGF}}
\label{model_params_pdgf} 
\begin{tabular}{p{1.5cm}p{7.8cm}p{3.25cm}p{2.1cm}p{0.75cm}}
\noalign{\hrule height 0.05cm}\noalign{\smallskip}
parameter & description & value & units & ref.  \\
\noalign{\smallskip}
 &  & [media, adventitia] &  &   \\
\noalign{\hrule height 0.05cm}\noalign{\smallskip}
 $D_{{}_P}$ & diffusivity & $0.1$ & [mm$^2$/day] & \citep{budu2008}\\[0.1cm]
  $r_{{}_{\eta}}$ & secretion ratio & [$0.5, 0.0$] & [-] & choice\\[0.1cm]
 $\eta_{{}_P}$ & autocrine secretion coefficient & $1.0 \times 10^{-6}$ & [mm$^3$/cell/day] & choice\\ [0.1cm]
    $l_{{}_{P1}}$ & drug-dose-dependent anti-inflammatory parameter  & $1.0$ & [nM]$^{-1}$ & choice\\ [0.1cm]
 $\varepsilon_{{}_P}$ & receptor internalization coefficient & $1.0 \times 10^{-7}$ & [mm$^3$/cell/day] & choice\\[0.1cm]
   $l_{{}_{P2}}$ & steepness coefficient for the drop in PDGF receptors  & $1.0 \times 10^{16}$ & [mm$^3$/mol] & choice\\ [0.1cm]
  $c_{{}_{T,th}}$ & TGF$-\beta$ threshold for the drop in PDGF receptors & $1.0 \times 10^{-16}$ & [mol/mm$^3$] & choice\\[0.1cm]
\noalign{\hrule height 0.05cm}\noalign{\smallskip}
\end{tabular}
\end{table}

\subsubsection{Transforming growth factor (TGF)-$\beta$ and extracellular matrix (ECM)}
Since the equations set up in \citep{MANJUNATHA2022106166} for the evolution of TGF-$\beta$ and ECM remain unchanged, we just list them down in their Eulerian forms below, which respectively read
\begin{eqnarray}
    \displaystyle{\left.\frac{\partial c_{{}_T}}{\partial t}\right|_{\bm{x}}} + \text{\sf div} \left(c_{{}_T}\,\boldsymbol{v}\right) = \underbrace{\text{\sf div} \left(D_{{}_{T}}\,\text{\sf grad}\,c_{{}_{T}}\right)}_{\text{diffusion}} - \underbrace{\overbrace{\varepsilon_{{}_T} \,\rho_{{}_{S}}\, c_{{}_{T}}}}^{=:\mathcal{S}_{{}_T}}_{\substack{\text{receptor}\\\text{internalization}}},\label{tgf_bal}\\
    \nonumber\\
    \displaystyle{\left.\frac{\partial c_{{}_C}}{\partial t}\right|_{\bm{x}}} + \text{\sf div} \left(c_{{}_C}\,\boldsymbol{v}\right) = \underbrace{\overbrace{\eta_{{}_C} \rho_{{}_{S}} \left(1 - \displaystyle{\frac{c_{{}_{C}}}{c_{{}_{C,th}}}}\right)}}^{=:\mathcal{R}_{{}_C}}_{\substack{\text{secretion by}\\\text{synthetic SMCs}}} - \underbrace{\overbrace{\varepsilon_{{}_C} \, c_{{}_{P}}\,c_{{}_{C}}}}^{=:\mathcal{S}_{{}_C}}_{\substack{\text{MMP-induced}\\\text{degradation}}},\label{ce_bal}
\end{eqnarray}
where $D_{{}_T}$ refers to the diffusivity of TGF-$\beta$ within the arterial wall, $\varepsilon_{{}_T}$ is termed the TGF-$\beta$ receptor internalization coefficient, $\eta_{{}_E}$ the collagen secretion coefficient, and $\varepsilon_{{}_E}$ the collagen degradation coefficient.

\begin{table}[htbp!]\
\centering
\caption{\textbf{Model parameters - TGF-$\beta$ and ECM}}
\label{model_params_tgf_ecm} 
\begin{tabular}{p{1.5cm}p{7.8cm}p{3.25cm}p{2.1cm}p{0.75cm}}
\noalign{\hrule height 0.05cm}\noalign{\smallskip}
parameter & description & value & units & ref.  \\[0.1cm]
\noalign{\hrule height 0.05cm}\noalign{\smallskip}
 \textbf{TGF-}$\bm{\beta}$ & & & \\[0.1cm]
 $D_{{}_T}$ & diffusivity& $0.1$ & [mm$^2$/day] & \citep{budu2008}\\
 $\varepsilon_{{}_T}$ & receptor internalization coefficient & $1.0 \times 10^{-7}$ & [mm$^3$/cell/day] & choice\\[0.2cm]
 \textbf{ECM} & & & \\[0.1cm]
 $\eta_{{}_C}$ & collagen secretion coefficient & $1.0 \times 10^{-8}$ & [mol/cell/day] & \citep{cilla2014}\\
 $\varepsilon_{{}_C}$ & collagen degradation coefficient & $1.0 \times 10^{21}$ & [mm$^3$/mol/day] & choice\\ 
 $c_{{}_{C,th}}$ & collagen secretion threshold & $7.0007 \times 10^{-9}$ & [mol/mm$^3$] & \citep{sae2013} \\[0.1cm]
\noalign{\hrule height 0.05cm}\noalign{\smallskip}
\end{tabular}
\end{table}

\subsubsection{Drug}
To incorporate the pharmacokinetics, the drug embedded within a polymer layer coated on the metallic core of the DES shall be introduced as an additional mediator $c_{{}_D}$ into the earlier developed modeling framework. Specifically, rapamycin-based drugs are intended to be investigated. The drug, since it is considered to be eluted from the stent surface, is devoid of any source terms within the arterial wall, rendering
\begin{equation}\label{drugsource}
    \mathcal{R}_{{}_{D}} := 0.
\end{equation}
A sink term takes care of the internalization of targeted receptors for the drug on the SMCs (FKBP12 for rapamycin-based drugs), which results in the deterioration of drug presence. This is formulated as a linear term depending on the local SMC density ($\rho_{{}_S}$) and the concentration of the drug itself ($c_{{}_D}$), and reads
\begin{equation}\label{drugsink}
    \mathcal{S}_{{}_{D}} := \varepsilon_{{}_{D1}}\,\rho_{{}_S}\,c_{{}_D},
\end{equation}
where $\varepsilon_{{}_{D1}}$ is termed the SMC-pertinent receptor internalisation coefficient for the drug.
\vspace{0.1in}\\
\noindent The drug is considered freely diffusive in the vessel wall, thereby being modeled by a standard diffusive term. Inserting Eqs. \ref{drugsource} and \ref{drugsink} into the general form in Eq. \ref{ard_eq_general_form}, the balance equation for the drug is then
\begin{equation}\label{drug_bal}
\displaystyle{\left.\frac{\partial c_{{}_D}}{\partial t}\right|_{\bm{x}}} + \text{\sf div} \left(c_{{}_D}\,\boldsymbol{v}\right) = \underbrace{\text{\sf div} \left(D_{{}_{D}}\,\text{\sf grad}\,c_{{}_{D}}\right)}_{\text{diffusion}} - \underbrace{\varepsilon_{{}_{D1}} \,\rho_{{}_{S}}\, c_{{}_{D}}}_{\substack{\text{receptor}\\\text{internalization}}},
\end{equation}
where $D_{{}_{D}}$ refers to the drug diffusivity. 
\vspace{0.1in}\\
\noindent One can also embed a more resolved drug transport model to capture the considered drug's receptor-specific and non-specific extracellular binding aspects (see for e.g. \cite{MCGINTY2015327,MCQUEEN2022992, salvi2022}) within the current framework with relative ease of implementation. 

\begin{table}[htbp!]\
\centering
\caption{\textbf{Model parameters - drug}}
\label{model_params_drug} 
\begin{tabular}{p{1.5cm}p{7.8cm}p{3.25cm}p{2.1cm}p{0.75cm}}
\noalign{\hrule height 0.05cm}\noalign{\smallskip}
parameter & description & value & units & ref.  \\[0.1cm]
\noalign{\hrule height 0.05cm}\noalign{\smallskip}
 $D_{{}_D}$ & diffusivity & $0.05 $ & [mm$^2$/day] & \citep{MCQUEEN2022992}\\[0.1cm]
 $\varepsilon_{{}_{D1}}$ & SMC receptor internalization coefficient & $1.0 \times 10^{-8}$ & [mm$^3$/cell/day] & choice\\[0.1cm]
\noalign{\hrule height 0.05cm}\noalign{\smallskip}
\end{tabular}
\end{table}

\subsubsection{Smooth muscle cells (SMCs)}
The modeling of SMCs remains largely unchanged from that in \citep{MANJUNATHA2022106166}, wherein the directional movement of SMCs are modeled via pseudoadvective chemotaxis and haptotaxis velocities given by
\begin{eqnarray}\label{vcvh}
    \bm{v}_{{}_{S1}} &:=& \chi_{{}_{S1}} \left(1 - \displaystyle{\frac{c_{{}_{C}}}{c_{{}_{C,th}}}}\right) \,\rho_{{}_{S}} \,\text{\sf grad}\,c_{{}_{P}}\nonumber\\
    \nonumber\\
    \bm{v}_{{}_{S2}} &:=& - \chi_{{}_{S2}} \,f_{{}_{S1}}(c_{{}_P}) \,\rho_{{}_{S}} \,\text{\sf grad}\,c_{{}_{C}},
\end{eqnarray}
wherein $\chi_{{}_{S1}}$ and $\chi_{{}_{S2}}$ denote the respective chemotactic and haptotactic sensitivities of the SMCs, and the sigmoidal scaling factor
\begin{equation}
        f_{{}_{S1}}(c_{{}_P}) := \frac{1}{1 + \exp\left({-l_{{}_{S1}}\left(c_{{}_P} - c_{{}_{P,th}}\right)}\right)}\,\,\,\in [0,1]
        \label{pdgf_scal}
\end{equation}
controls the switching on of the haptotactic SMC movement beyond a certain threshold of PDGF concentration $c_{{}_{P,th}}$,  $l_{{}_{S1}}$ modulating the steepness of this switch.
\vspace{0.1in}\\
\noindent The SMC proliferation assays conducted by \citet{koyama1994} indicate that the SMC proliferation saturates beyond a certain level of PDGF stimulation. In this regard, we forego a linear dependence of the proliferation of SMCs on PDGF concentration ($c_{{}_P}$) \citep{MANJUNATHA2022106166} and introduce a function 
\begin{equation}\label{fP_prolif}
    f_{{}_{S2}}(c_{{}_P}) := 1 - \exp(-l_{{}_{S2}}\,c_{{}_P}) \in [0,1],
\end{equation}
where $l_{{}_{S2}}$ is termed the PDGF receptor saturation coefficient. 

\begin{figure}[htbp]
    \centering
    \includegraphics[scale=0.6]{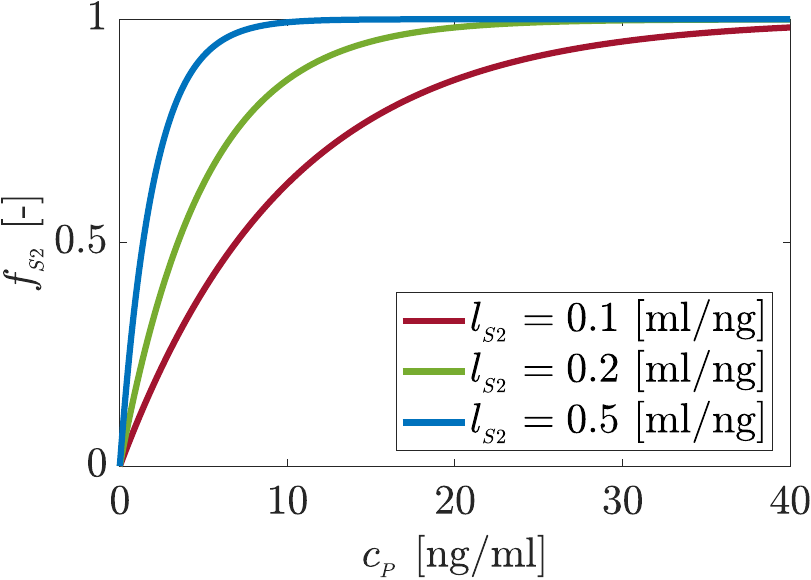}
    \caption{\textbf{PDGF-dependent SMC proliferation scaling function}}
    \label{fig_fS2}
\end{figure}

\noindent The TGF$-\beta$ induced modulation of SMC proliferation is taken into account as in \citep{MANJUNATHA2022106166} via a scaling function dependent on the TGF$-\beta$ concentration ($c_{{}_T}$) which is given by
    \begin{equation}
        f_{{}_{S3}}(c_{{}_T}) := \frac{1}{1 + \exp\left({l_{{}_{S3}}\left(c_{{}_T} - c_{{}_{T,th}}\right)}\right)}\,\,\,\in [0,1],
        \label{tgf_scal_SMC}
    \end{equation}
where $l_{{}_{S3}}$ controls steepness of this modulatory effect, and $c_{{}_{T,th}}$ is the related TGF$-\beta$ concentration threshold.
\vspace{0.1in}\\
\noindent To incorporate the direct dose-dependent effect of rapamycin-based drugs on the inhibition of proliferation, an additional drug-dependent scaling function $f_{{}_{S4}}(c_{{}_D})$ has to be introduced. Based on cell proliferation assays conducted by \citet{PARRY200638}, they suggest the Hill type of functions to accommodate the aforementioned effect (sirolimus in this case). We hence incorporate the complimentary curve in our model to scale the proliferativity of SMCs, which reads
\begin{equation}\label{fDs}
    f_{{}_{S4}}(c_{{}_D}) := 1 - \frac{1}{100}\,\left(\frac{A_{{}_S}\,c^{\alpha}_{{}_D}}{c^{\alpha}_{{}_D} + B_{{}_S}^{\alpha}}\right),
\end{equation}
where $A_{{}_S}$ is termed the maximum efficacy of the drug in inhibition of SMC proliferation, $B_{{}_S}$ the drug dose that results in half efficacy, and $\alpha$ the Hill coefficient, $B_{{}_S}$ and $c_{{}_D}$ being expressed in [nM] units.
\vspace{0.1in}\\
The source term for SMCs is hence arrived at to be of the form
\begin{equation}\label{smcprolif}
    \mathcal{R}_{{}_{S}} := \eta_{{}_S}\, f_{{}_{S2}}(c_{{}_P})\,f_{{}_{S3}}(c_{{}_T})\,f_{{}_{S4}}(c_{{}_D})\, \rho_{{}_{S}} \left(1 - \displaystyle{\frac{c_{{}_{C}}}{c_{{}_{C,th}}}}\right),
\end{equation}
where $\eta_{{}_S}$ is the SMC proliferation coefficient. The constant apoptosis is not intended to be modeled as part of this framework. Hence,
\begin{equation}\label{sink_SMC}
    \mathcal{S}_{{}_S} := 0.
\end{equation}
\noindent Using Eqs. \ref{ard_eq_general_form}, \ref{vcvh}, \ref{sink_SMC}, and \ref{smcprolif}, the particularized governing equation for the SMC density is then arrived at to be
\begin{equation}\label{smc_bal}
\begin{aligned}
    \displaystyle{\left.\frac{\partial \rho_{{}_S}}{\partial t}\right|_{\bm{x}}} + \text{\sf div} \left(\rho_{{}_S}\,\boldsymbol{v}\right) = 
    & - \underbrace{\text{\sf div}\left(\rho_{{}_S}\,\overbrace{\chi_{{}_{S1}} \left(1 - \displaystyle{\frac{c_{{}_{C}}}{c_{{}_{C,th}}}}\right)\,\text{\sf grad}\,c_{{}_{P}}}^{ \bm{v}_{{}_{S1}}} \right)}_{\text{chemotaxis}} \\
    & + \underbrace{ \text{\sf div}\left(\rho_{{}_S}\,\overbrace{\chi_{{}_{S2}} \,f_{{}_{S1}}(c_{{}_P}) \,\text{\sf grad}\,c_{{}_{C}}}^{- \bm{v}_{{}_{S2}}}\right)}_{\text{haptotaxis}}\\
    & + \underbrace{\eta_{{}_S}\, f_{{}_{S2}}(c_{{}_P})\,f_{{}_{S3}}(c_{{}_T})\,f_{{}_{S4}}(c_{{}_D})\, \rho_{{}_{S}} \left(1 - \displaystyle{\frac{c_{{}_{C}}}{c_{{}_{C,th}}}}\right)}_{\text{proliferation}}.
\end{aligned}
\end{equation}

\begin{table}[htbp!]\
\centering
\caption{\textbf{Model parameters - SMCs}}
\label{model_params_SMC} 
\begin{tabular}{p{1.5cm}p{6.8cm}p{3.25cm}p{2.1cm}p{0.75cm}}
\noalign{\hrule height 0.05cm}\noalign{\smallskip}
parameter & description & value & units & ref.  \\[0.1cm]
\noalign{\smallskip}
 &  & [media, adventitia] &  & \\
\noalign{\hrule height 0.05cm}\noalign{\smallskip}
 $\chi_{{}_{S1}}$ & chemotactic sensitivity & [$1.0 \times 10^{14}, 1.0 \times 10^{13}$] & [mm$^5$/mol/day] & \citep{budu2008}\\ [0.1cm]
 $\chi_{{}_{S2}}$ & haptotactic sensitivity & [$1.0 \times 10^{7}, 1.0 \times 10^{6}$] & [mm$^5$/mol/day] & choice\\ [0.1cm]
 $l_{{}_{S1}}$ & steepness coefficient for 
haptotaxis activation & $1.0 \times 10^{16}$ & [mm$^{3}$/mol] & choice\\ [0.1cm]
$c_{{}_{P,th}}$ & PDGF threshold for 
haptotaxis activation & $1.0 \times 10^{-15}$ & [mol/mm$^3$] & choice \\ [0.1cm] 
 $\eta_{{}_S}$ & proliferation coefficient & [$0.4, 0.2$] & [day]$^{-1}$ & choice\\ [0.1cm]
 $l_{{}_{S2}}$ & PDGF receptor saturation coefficient
& $1.0 \times 10^{-7}$ & [ml/ng] & \citep{koyama1994}\\ [0.1cm]
 $l_{{}_{S3}}$ & steepness coefficient for 
TGF$-\beta$-induced proliferation modulation
& $1.0 \times 10^{16}$ & [mm$^3$/mol] & choice\\ [0.6cm]
$c_{{}_{T,th}}$ & TGF$-\beta$ threshold for proliferation inhibition & $1.0 \times 10^{-16}$ & [mol/mm$^3$] & choice\\[0.1cm]
$A_{{}_S}$ & maximum efficacy of the drug
against $\quad \quad$ SMC proliferation & $69.8$ & [$\%$] & \citep{PARRY200638}\\[0.6cm]
$B_{{}_S}$ & drug concentration for half efficacy & $2.2$ & [nM] & \citep{PARRY200638}\\[0.1cm]
$\alpha$ & Hill coefficient for drug-induced $\quad \quad \quad \quad$ SMC proliferation inhibition & $2.96$ & [ - ] & \citep{PARRY200638}\\[0.6cm]
\noalign{\hrule height 0.05cm}\noalign{\smallskip}
\end{tabular}
\end{table}

\subsubsection{Endothelial cells (ECs)}
The endothelium is a significant mediator for the restenotic process since it moderates the platelet activity as well as the inflammatory response of the arterial wall after stent implantation. The current work intends to incorporate the spatiotemporal effects of stent-induced endothelial denudation and pharmacodynamics-modulated endothelial healing. In this regard, we introduce an additional variable for the EC density $\rho_{{}_{E}}$ that resides on the luminal surface only, since the endothelium is a monolayer of ECs. 

\myparagraph{EC proliferation}
ECs are assumed to proliferate along the endothelium in a logistic fashion. A source term is thereby introduced which looks very similar to the one for ECM secretion in Eq. \ref{ce_bal} but with an additional drug-dependent function $f_{{}_{E1}}$ appended to it, which therefore takes the form
\begin{equation}\label{ensource}
    \mathcal{R}_{{}_{E}} := \eta_{{}_{E}}\,f_{{}_{E1}}(c_{{}_D})\,\rho_{{}_{E}}\,\left(1 - \frac{\rho_{{}_{E}}}{\rho_{{}_{E,eq}}}\right),
\end{equation}
where $\eta_{{}_{E}}$ is termed the EC proliferation coefficient, and $\rho_{{}_{E,eq}}$ is the EC density on a healthy endothelium in homeostasis. For what form of $f_{{}_{E1}}(c_{{}_D})$ to embed, we again turn our attention to a complementary curve to a Hill-type function as in Eq. \ref{fDs}, i.e.,
\begin{equation}\label{fDenP}
    f_{{}_{E1}}(c_{{}_D}) := 1 - \frac{1}{100}\,\left(\frac{A_{{}_E}\,c^{\beta}_{{}_D}}{c^{\beta}_{{}_D} + B_{{}_E}^{\beta}}\right),
\end{equation}
where the constants $A_{{}_E}$, $B_{{}_E}$, and $\beta$ can be fit to data (see for e.q. \citep{barilli2008}). 

\begin{figure}[htbp]
    \centering
    \includegraphics[scale=0.6]{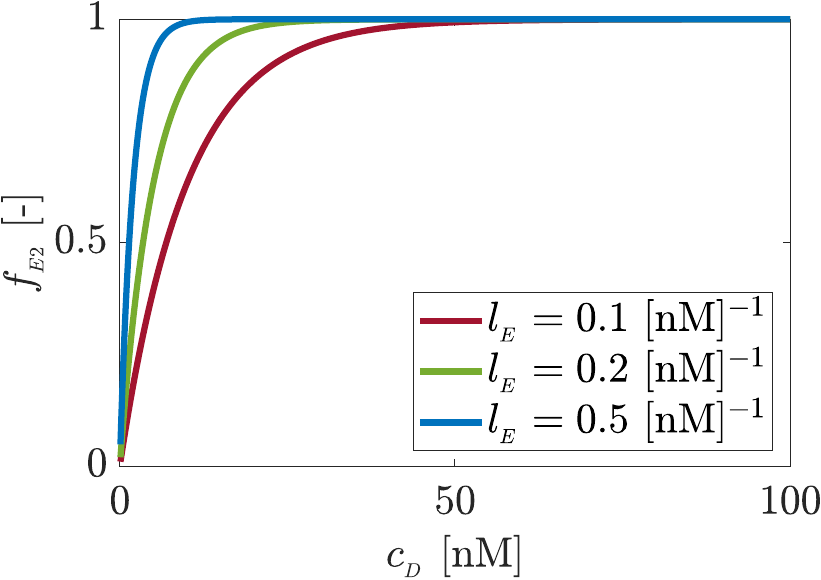}
    \caption{\textbf{Drug-dependent scaling function for the apoptosis of ECs}}
    \label{fig_fE2_cD}
\end{figure}

\myparagraph{EC apoptosis}
Although in a homeostatic environment, apoptosis, which is a highly controlled death of cells in multicellular organisms, is a natural part of the turnover of cellular constituents, in an inflammatory setting,  an imbalance persists in the quantitative levels of apoptosis and replenishment of cells. In the case of ECs, this disparity is known to have been exacerbated by the presence of rapamycin-based drugs embedded in DESs \citep{barilli2008}. To model this effect, we introduce herein a sink term of the form
\begin{equation}\label{ensink}
    \mathcal{S}_{{}_{E}} := \varepsilon_{{}_{E}}\,f_{{}_{E2}}(c_{{}_D})\,\rho_{{}_{E}},
\end{equation}
where $\varepsilon_{{}_{E}}$ is termed the EC apoptosis coefficient, and $f_{{}_{E2}}(c_{{}_D})$, being termed the drug-dependent apoptosis function chosen to be of the form as in Eq. \ref{fP_prolif}, i.e.,
\begin{equation}
    f_{{}_{E2}}(c_{{}D}) := 1 - \exp(-l_{{}_E}\,c_{{}_D}) \in [0, 1].
\end{equation}
Here, $l_{{}_E}$ controls the dose-dependent exacerbation of apoptosis (see Fig. \ref{fig_fE2_cD}).
\paragraph{Remark} The above modeling terms are not sufficient enough to capture the behavior of the ECs, since there shall be no regeneration of the endothelium where the initial value of the EC density ($\rho_{{}_E}$) is prescribed to be zero on the assumption of complete denudation ($\mathcal{R}_{{}_E} = 0$ when $\rho_{{}_E} = 0$). If one chooses the source term to just be of the form
\begin{equation}\label{ensource_mod}
    \mathcal{R}_{{}_{E}} := \eta_{{}_{E}}\,f_{{}_{E1}}(c_{{}_D})\,\left(1 - \frac{\rho_{{}_{E}}}{\rho_{{}_{E,eq}}}\right),
\end{equation}
then, in the absence of the drug, the endothelial recovery happens homogenously throughout the denuded region. This is deemed not to be physiological, and endothelial regeneration is observed to occur from edges inward. To this end, assuming a minuscule amount of diffusivity of ECs shall enable modeling this process more accurately.
\vspace{0.1in}\\
\noindent The flow-dependent and chemotactic movements of the ECs are neglected in this work. Considering the ECs to be freely diffusive on the luminal side of the arterial wall, and using Eqs. \ref{ard_eq_general_form}, \ref{ensource}, and \ref{ensink}, the particularized differential equation for the ECs reads
\begin{equation}
\displaystyle{\left.\frac{\partial \rho_{{}_{E}}}{\partial t}\right|_{\bm{x}}} + \text{\sf div}_{{}_{\Gamma}} \left(\rho_{{}_{E}}\,\boldsymbol{v}_{{}_{\Gamma}}\right) = \text{\sf div}_{{}_{\Gamma}} \left(D_{{}_{E}}\,\text{\sf grad}_{{}_{\Gamma}}\,\rho_{{}_{E}}\right)
+ \underbrace{\eta_{{}_{E}}\, f_{{}_{E1}}(c_{{}_D})\,\rho_{{}_{E}}\,\left(1 - \frac{\rho_{{}_{E}}}{\rho_{{}_{E,eq}}}\right)}_{\text{proliferation}} - \underbrace{\varepsilon_{{}_{E}}\,f_{{}_{E2}}(c_{{}_D})\,\rho_{{}_{E}}}_{\text{apoptosis}}
\end{equation} 
where $D_{{}_{E}}$ is the diffusivity of ECs, $\text{\sf div}_{{}_{\Gamma}}$ and $\text{\sf grad}_{{}_{\Gamma}}$ referring to surface divergence and gradient operators respectively. 

\begin{table}[htbp!]\
\centering
\caption{\textbf{Model parameters - ECs}}
\label{model_params_EC} 
\begin{tabular}{p{1.5cm}p{6.8cm}p{3.25cm}p{2.1cm}p{0.75cm}}
\noalign{\hrule height 0.05cm}\noalign{\smallskip}
parameter & description & value & units & ref.  \\[0.1cm]
\noalign{\hrule height 0.05cm}\noalign{\smallskip}

$D_{{}_E}$ & diffusivity & $0.01$ & [mm$^2$/day] & choice\\[0.1cm] 
 $\eta_{{}_E}$ & proliferation coefficient & $0.1$ & [day]$^{-1}$ & choice\\ [0.1cm]
 $\rho_{{}_{E,eq}}$ & EC density on healthy vascular endothelia & $500$ & [cells/mm$^2$] & \citep{napione2012} \\ [0.1cm] 
$A_{{}_E}$ & maximum efficacy of the drug
against $\quad \quad$ EC proliferation & $65.4$ & [$\%$] & \citep{PARRY200519}\\[0.6cm]
$B_{{}_E}$ & drug concentration for half efficacy & $9.19$ & [nM] & \citep{PARRY200519}\\[0.1cm]
$\beta$ & Hill coefficient for drug-induced $\quad \quad \quad \quad$ EC proliferation inhibition & $1.46$ & [ - ] & \citep{PARRY200519}\\[0.6cm]
 $\varepsilon_{{}_E}$ & apoptosis coefficient & $0.01$ & [day]$^{-1}$ & choice\\ [0.1cm]
  $l_{{}_E}$ & drug-dose-dependent apoptosis parameter & $0.2$ & [nM]$^{-1}$ & \citep{barilli2008}\\ [0.1cm]
\noalign{\hrule height 0.05cm}\noalign{\smallskip}
\end{tabular}
\end{table}

\subsection{Continuum mechanical modeling}
Arterial media and adventitia are considered to be the key layers influencing the structural behavior of the vessel wall, each of which is assumed to be composed of two families of collagen fibres embedded in an isotropic ground matrix. Proliferativity of the SMCs drives the growth process within the isotropic ground matrix. Collagen, and hence the extracellular matrix, is assumed to modulate the compliance of the arterial wall.   

\subsubsection{Kinematics}\label{kinematics_section}
If $\boldsymbol{\varphi}$ is the deformation map between the reference configuration $\Omega_0$ at time $t_0$ and the current configuration $\Omega$ at time $t$ of a continuum body, a particle at position $\boldsymbol{X}$ in the reference configuration is mapped to that at $\boldsymbol{x}$ in the current configuration via the deformation gradient $\boldsymbol{F} = \partial\,\boldsymbol{\varphi}(\boldsymbol{X},t)/\partial \bm{X}$. The right and the left Cauchy-Green tensors are further defined by $\boldsymbol{C} = \boldsymbol{F}^T \,\boldsymbol{F} $ and $\boldsymbol{B} = \boldsymbol{F} \,\boldsymbol{F}^T $, respectively.

For the description of growth, the well-established multiplicative decomposition of the deformation gradient \citep{rodriguez1994} is adopted, i.e.
\begin{equation}
    \boldsymbol{F} = \boldsymbol{F}_e\,\boldsymbol{F}_g,
\end{equation}
wherein an incompatible intermediate configuration that achieves a locally stress-free state is achieved via the mapping $\bm{F}_g$, and the elastic deformation gradient $\bm{F}_e$ ensures the compatibility of the deformations in the continuum. The elastic right Cauchy-Green tensor and the growth-related right and left Cauchy-Green tensors hence take the forms
\begin{eqnarray}\label{ce_cg_bg}
    \bm{C}_e &:=& \bm{F}^T_e\,\bm{F}_e = \bm{F}^{-T}_g\,\bm{C}\,\bm{F}^{-1}_g \nonumber\\
    \bm{C}_g &:=& \bm{F}^T_g\,\bm{F}_g\\
    \bm{B}_g &:=& \bm{F}_g\,\bm{F}^T_g.\nonumber
\end{eqnarray}

\subsubsection{Structural tensors}\label{struct_tensors}

Since the arterial wall is assumed to be composed of collagen fibres arranged in two distinct helices, the notion of structural tensors is introduced in this context to model the initial anisotropic behavior. If $\bm{a}_{01}$ and $\bm{a}_{02}$ represent the two local collagen orientations in the reference configuration, the associated structural tensors are defined to be
\begin{eqnarray}\label{struct_tensors_def}
    \bm{H}_1 : = \bm{a}_{01}\otimes\bm{a}_{01},\quad \bm{H}_2 := \bm{a}_{02}\otimes\bm{a}_{02}.
\end{eqnarray}
Owing to the symmetry of the above definition, the mapping of the respective structural tensors to the intermediate configuration is achieved by
\begin{equation}
    \hat{\bm{H}}_i = \bm{F}_g\,\bm{H}_i\,\bm{F}^T_g, \quad i = 1,2,
\end{equation}
quantities residing in the intermediate configuration being depicted in the form $\hat{(\bullet)}$.

\subsubsection{Helmholtz free energy}
The Helmholtz free energy per unit volume in the reference configuration $\Omega_0$ is split into an isotropic part associated with the isotropic ground matrix, and an anisotropic part corresponding to the collagen fibres. Employing the idea of material isomorphism in the context of inelasticity \citep{SVENDSEN1998473, reese2021}, the Helmholtz free energy per unit reference volume is defined to be an isotropic function of the selected set of arguments that reside in the stress-free intermediate configuration as
\begin{equation}
\psi := \bar{\psi}(\bm{C}_e, \bm{B}_g, \hat{\bm{H}}_1, \hat{\bm{H}}_2, \hat{c}_{{}_E}).  \label{hfe}
\end{equation} 
As elucidated in \citep{HOLTHUSEN2023105174}, the definition of this stress-free intermediate configuration suffers from the lack of rotational uniqueness. In this context, \citet{HOLTHUSEN2023105174} suggest performing a pull-back of the kinematic quantities and the structural tensors defined in Eqs. \ref{ce_cg_bg} and \ref{struct_tensors_def} to the so-called co-rotated intermediate configuration by using the rotational part $\bm{R}_g \in \mathcal{SO}(3)$  of the growth-related deformation gradient $\bm{F}_g = \bm{R}_g\,\bm{U}_g$, i.e.,
\begin{eqnarray}\label{kinematicquantities}
    \bar{\bm{C}}_e &:=& \bm{R}^{-1}_g\,\bm{C}_e\,\bm{R}_g = \bm{U}^{-1}_g\,\bm{C}\,\bm{U}^{-1}_g\nonumber\\
    \nonumber\\
    \bar{\bm{B}}_g &:=& \bm{R}^{-1}_g\,\bm{B}_g\,\bm{R}_g \nonumber\\
    \nonumber\\
    \bar{\bm{H}}_i &:=& \bm{R}^{-1}_g\,\hat{\bm{H}}_i\,\bm{R}_g = \bm{U}_g\,\bm{H}\,\bm{U}_g.
\end{eqnarray}
The volumetric change associated with the deformation gradient $\bm{F}$ is deduced to be
\begin{equation}\label{detFsplit}
    J = \text{\sf det}\,\bm{F} = \bar{J}_e\,J_g, \quad \bar{J}_e = \sqrt{\det \bm{C}_e}, \quad J_g = \det \bm{U}_g.
\end{equation}
Quantities residing in the aforementioned co-rotated intermediate configuration are depicted in the form $\bar{(\bullet)}$. Due to the fact that the Helmholtz free energy is defined to be an isotropic function of its arguments, and that $\bm{C}_g$ and $\bm{B}_g$ are in possession of the same eigenvalues, and additionally, due to the invariance of the eigenvalues with respect to rotations, the argument $\bm{B}_g$ in Eq. \ref{hfe} can be replaced by $\bm{C}_g$. Hence the referential Helmholtz free energy per unit volume can alternatively be expressed as
\begin{equation}
    \psi := \bar{\psi}(\bar{\bm{C}}_e, \bm{C}_g, \bar{\bm{H}}_1, \bar{\bm{H}}_2, \bar{c}_{{}_C}).  \label{hfe_alt1}
\end{equation}

\subsubsection{Clausius-Duhem inequality}
The arterial wall and most soft biological tissues are open systems, allowing for heat and mass exchange with the surroundings. In this regard, we consider the isothermal form of the Clausius-Duhem/dissipation inequality suggested by \citet{Kuhl2003TheoryAN} for deriving further insights into the growth process which reads
\begin{equation}\label{cde1}
    -\dot{\psi} + \frac{1}{2}\bm{S}:\dot{\bm{C}} + \mathcal{R}_0 \geq 0,
\end{equation}
where $\bm{S}$ is the second Piola-Kirchhoff stress tensor, while $\mathcal{R}_0$ encompasses the entropy sources, sinks, and fluxes that account for biochemical and cellular interactions. The material rate of Helmholtz free energy hence is given by
\begin{equation}\label{psidot}
    \dot{\psi} = \dot{\bar{\psi}} = \frac{\partial \bar{\psi}}{\partial \bar{\bm{C}}_e}:\dot{\bar{\bm{C}}}_e + \frac{\partial \bar{\psi}}{\partial \bm{C}_g}\,\dot{\bm{C}}_g + \frac{\partial \bar{\psi}}{\partial \bar{\bm{H}}_1}:\dot{\bar{\bm{H}}}_1 + \frac{\partial \bar{\psi}}{\partial \bar{\bm{H}}_2}:\dot{\bar{\bm{H}}}_2 + \frac{\partial \bar{\psi}}{\partial \bar{c}_{{}_C}}:\dot{\bar{c}}_{{}_C},
\end{equation}
with
\begin{eqnarray}\label{cedot}
    \dot{\bar{\bm{C}}}_e  = \bm{U}^{-1}_g\,\dot{\bm{C}}\,\bm{U}^{-1}_g - \bar{\bm{C}}_e\,\bar{\bm{L}}_g - \bar{\bm{L}}^T_g\,\bar{\bm{C}}_e,
\end{eqnarray}
where
\begin{equation}\label{lgbar}
    \bar{\bm{L}}_g := \dot{\bm{U}}_g\,\bm{U}^{-1}_g.
\end{equation}
Further, it can be shown that
\begin{equation}\label{cgdot}
    \dot{\bm{C}}_g = \dot{\bm{U}}_g\,\bm{U}_g + \bm{U}_g\,\dot{\bm{U}}_g = \bar{\bm{L}}_g\,\bm{C}_g + \bm{C}_g\,\bar{\bm{L}}^T_g
\end{equation}
and
\begin{equation}\label{hbardot}
    \dot{\bar{\bm{H}}}_i = \bm{U}_g\,\dot{\bm{H}}_i\,\bm{U}_g + \bar{\bm{L}}_g\,\bar{\bm{H}}_i + \bar{\bm{H}}_i\,\bar{\bm{L}}^T_g,\quad i =1,2. 
\end{equation}
In addition, the ECM concentration in the co-rotated intermediate configuration is $\bar{c}_{{}_C} = J^{-1}_g\,c^0_{{}_C}$, which leads to
\begin{eqnarray}\label{cebardot}
    \frac{\partial \bar{\psi}}{\partial \bar{c}_{{}_C}} &=& J_g\,\frac{\partial \bar{\psi}}{\partial c^0_{{}_C}}\nonumber\\
    \nonumber\\
    \dot{\bar{c}}_{{}_C} &=& J^{-1}_g\,\left(\dot{c}^0_{{}_C} - c^0_{{}_C}\,\bm{I}:\bm{\bar{L}}_g\right)
\end{eqnarray}
since $\dot{J}_g = J_g\,{\sf tr} (\bm{\bar{L}}_g)$.
Plugging Eqs. \ref{cedot} to \ref{cebardot} into the Clausius-Duhem inequality and employing the Colemann-Noll procedure \citep{coleman_noll1963} shall result in the definition of the second Piola-Kirchhoff stress tensor to be
\begin{equation}
    \bm{S} = 2\,\bm{U}^{-1}_g\,\frac{\partial \bar{\psi}}{\partial \bar{\bm{C}}_e}\,\bm{U}^{-1}_g.
\end{equation}
The Clausius-Duhem inequality hence attains the reduced format
\begin{equation}\label{cde_1}
    \mathscr{D}_{{}_{\sf red}} := \left[\bar{\bm{\Sigma}} - \bar{\bm{X}} - \bar{\bm{\Upsilon}} + c^{0}_{{}_C}\,\frac{\partial \bar{\psi}}{\partial c^0_{{}_C}}\,\bm{I}\right]:\bar{\bm{L}}_g - \bm{G}_1:\dot{\bm{H}}_1 - \bm{G}_2:\dot{\bm{H}}_2 - \frac{\partial \bar{\psi}}{\partial {c}^0_{{}_C}}\,\dot{{c}}^0_{{}_C} + \mathcal{R}_0 \geq 0,
\end{equation}
where $\bar{\bm{\Sigma}}$ and $\bar{\bm{X}}$ are the equivalents of the so-called Mandel and back stress tensors, respectively, in the co-rotated intermediate configuration. See \ref{cde} for the definitions of the other tensors involved in the above expression.

\subsubsection{Volumetric growth}\label{vg}
It is here that we introduce the definition for the stretch part $\bm{U}_g$ of the growth-related deformation gradient $\bm{U}_g$ to be of the form prescribed in \citep{MANJUNATHA2022106166}, in line with the suggestion by \citet{Lubarda2002OnTM} for transversely isotropic growth, i.e.,
\begin{equation}\label{ug}
    \bm{U}_g := \bm{I} + (\vartheta - 1)\,\bm{\gamma}\otimes\bm{\gamma},
\end{equation}
where
\begin{equation}
    \bm{\gamma} := \frac{\bm{a}_{01} \times \bm{a}_{02}}{||\bm{a}_{01} \times \bm{a}_{02}||}.
\end{equation}
The physical interpretation behind the above definition is that if the growth manifests in a direction orthogonal to the plane consisting of the collagen fibres, the intermediate grown configuration can be deemed stress-free, and so can the previously defined co-rotated intermediate configuration. 
With the local referential SMC density $\rho^0_{{}_S} = J^{-1}\,\rho_{{}_S}$ obtained from the balance equation for SMCs (Eq. \ref{smc_bal}) in hand, and assuming the growth process to arrange the SMCs in such a way that 
\begin{equation}
\hat{\rho}_{{}_S} = \bar{\rho}_{{}_S} = {\rho}_{{}_{S,eq}},
\end{equation}
where $\rho_{{}_{S,eq}}$ refers to the SMC density in a healthy homeostatic artery, we can define the volumetric change associated with growth to be
\begin{equation}
    J_{g} = \vartheta := \frac{\rho^0_{{}_S}}{\rho_{{}_{S,eq}}}.
\end{equation}
Within the context of this work, we forego the concept of remodeling within the arterial wall since in-stent restenosis results from an inflammatory response wherein growth is adjudged to be the dominant effect compared to remodeling. Pathophysiologies including hypertension, fibrosis, aortic aneurysm, etc. warrant considering the remodeling aspects in much detail. Therein, the concept of tensional homeostasis can be exploited \cite{HOLTHUSEN2023105174, lamm2022}. It is therefore assumed that the referential structural tensors $\bm{H}_1$ and $\bm{H}_2$ do not evolve in time rendering $\dot{\bm{H}}_1 = \dot{\bm{H}}_2 = \bm{0}$. The reduced dissipation inequality hence boils down (see \ref{cde}) to the form
\begin{equation}\label{red_cde}
        \mathscr{D}_{{}_{\sf red}} := \left[\bar{\bm{\Sigma}} - \bar{\bm{X}} + c^{0}_{{}_C}\,\frac{\partial \bar{\psi}}{\partial c^0_{{}_C}}\,\bm{I}\right]:\bar{\bm{L}}_g - \frac{\partial \bar{\psi}}{\partial {c}^0_{{}_C}}\,\dot{{c}}^0_{{}_C} + \mathcal{R}_0 \geq 0.
\end{equation}
Due to $\bm{U}_g$, and hence $\bar{\bm{L}}_g$ (Eq. \ref{lgbar}), being prescribed, the expressions for the back stress -like tensor $\bar{\bm{X}}$ and the Mandel-like stress tensor $\bar{\bm{\Sigma}}$ will be an outcome of the dissipation inequality leading to
\begin{equation}
    \bar{\bm{\Sigma}} = \bar{\bm{X}} - c^{0}_{{}_C}\,\frac{\partial \bar{\psi}}{\partial c^0_{{}_C}}\,\bm{I},  \quad \bar{\bm{X}} = 2\,\vartheta^2\,\frac{\partial \bar{\psi}}{\partial \bm{C}_g},
\end{equation}
further condensing the dissipation inequality to
\begin{equation}
    \mathscr{D}_{{}_{\sf red}} := - \frac{\partial \bar{\psi}}{\partial {c}^0_{{}_C}}\,\dot{{c}}^0_{{}_C} + \mathcal{R}_0 \geq 0.
\end{equation}
The above inequality sheds light on the fact that when $\dot{c}^0_{{}_E} \leq 0$, i.e., a degradation in collagen content, it represents a thermodynamically consistent growth process without the entropy source, i.e., $\mathcal{R}_0 = 0$. But when $\dot{c}^0_{{}_E}>0$, which represents secretion of collagen, and hence induction of local order in the cellular arrangements, the entropy source has to be non-zero locally, specifically, 
\begin{equation}
    \mathcal{R}_0 \geq \frac{\partial \bar{\psi}}{\partial {c}^0_{{}_C}}\,\dot{{c}}^0_{{}_C},
\end{equation}
the arterial wall being an open system allowing for this to be the case.
\subsubsection{Choices for the Helmholtz free energies}
The specific choice for the non-collagenous isotropic ground matrix, consisting of elastin and SMCs, is assumed to be of the Neo-Hookean form, given by
\begin{equation}
\psi_{iso} = \displaystyle{\frac{\mu}{2}}\left(\text{tr}\,\bar{\boldsymbol{C}}_e - 3\right) - \mu\,\text{ln}\,\bar{J}_e + \displaystyle{\frac{\Lambda}{4}}\left(\bar{J}_e^2 - 1 - 2\,\text{ln}\,\bar{J}_e \right),
\label{iso_hfe}
\end{equation} 
where the definition of $\bar{\bm{C}}_e$ from Eq. \ref{kinematicquantities}, and that of $\bar{J}_e$ from Eq. \ref{detFsplit} are utilized. The anisotropic part modeling the behaviour of collagen is particularized to be of exponential form \citep{holzapfel2000} as
\begin{equation}
\psi_{ani} = \displaystyle{\frac{k_1}{2k_2}}\sum_{i=1,2} \left(\text{\sf exp}\left[k_2\langle \bar{E}_i\rangle ^2 \right]-1\right),
\label{aniso_hfe}
\end{equation}
where $\bar{E}_{i} = \bm{H}_i:\bar{\bm{C}}_e - 1,\,\, i = 1,2$ is the Green-Lagrange strain in the direction of the collagen fibres, and the Macaulay brackets ensure that the fibres are activated only in tension. The stress-like material parameter $k_1$ in the above definition is here considered to be a linear function of the local ECM concentration in the co-rotated intermediate configuration $\bar{c}_{{}_E}$, i.e.,
\begin{equation}
    k_1 := \bar{k}_1 \,\displaystyle{\left(\frac{\bar{c}_{{}_C}}{c_{{}_{C,eq}}}\right)},
\end{equation}
$\bar{k}_1$ being the stress-like material parameter for healthy collagen, and $c_{{}_{E,eq}}$ referring to the homeostatic ECM concentration in a healthy artery. The total Helmholtz free energy per unit referential volume is then the summation
\begin{equation}
    \psi = \psi_{iso} + \psi_{ani}.
\end{equation}

\subsubsection{Balance of linear momentum}
Since the restenotic process occurs over a period of months, the growth is considered quasistatic, thereby rendering the effects of inertia of the added masses negligible. Hence the quasistatic form of the balance of linear momentum is enforced on the arterial wall, given by
\begin{equation}\label{mom_bal}
{\sf Div}\, (\bm{F}\,\bm{S}) + \boldsymbol{B} = \boldsymbol{0},
\end{equation}
where $\bm{B}$ is the body force vector.

\begin{table}[htbp!]\
\centering
\caption{\textbf{Model parameters - structural}}
\label{model_params_struct} 
\begin{tabular}{p{1.5cm}p{7.25cm}p{3.25cm}p{2.1cm}p{0.75cm}}
\noalign{\hrule height 0.05cm}\noalign{\smallskip}
parameter & description & value & units & ref.  \\
\noalign{\smallskip}
 &  & [media, adventitia] &  &   \\
\noalign{\hrule height 0.05cm}\noalign{\smallskip}
 $\mu$ & shear modulus for the matrix & [$0.02, 0.008$] & [M Pa] & \citep{he2020}\\[0.1cm]
 $\Lambda$ & Lam\'{e} parameter for the matrix & $10$ & [M Pa] & \citep{he2020}\\[0.1cm]
 $\bar{k}_1$ & stress-like parameter for collagen fibres & [$0.112, 0.362$] & [M Pa] & \citep{he2020}\\ [0.1cm]
 $k_2$ & exponential coefficient for collagen fibres & [$20.61, 7.089$] & [-] & \citep{he2020}\\ [0.1cm]
 $\alpha_{{}_a}$ & collagen orientation angle w.r.t circumference & [$41, 50.1$] & [${}^{\circ}$] & \citep{he2020}\\ [0.1cm]
\noalign{\hrule height 0.05cm}\noalign{\smallskip}
\end{tabular}
\end{table}

\newpage
\subsection{Boundary conditions}\label{bc}

\begin{figure}[htbp!]
    \centering
    \includegraphics[scale=0.8]{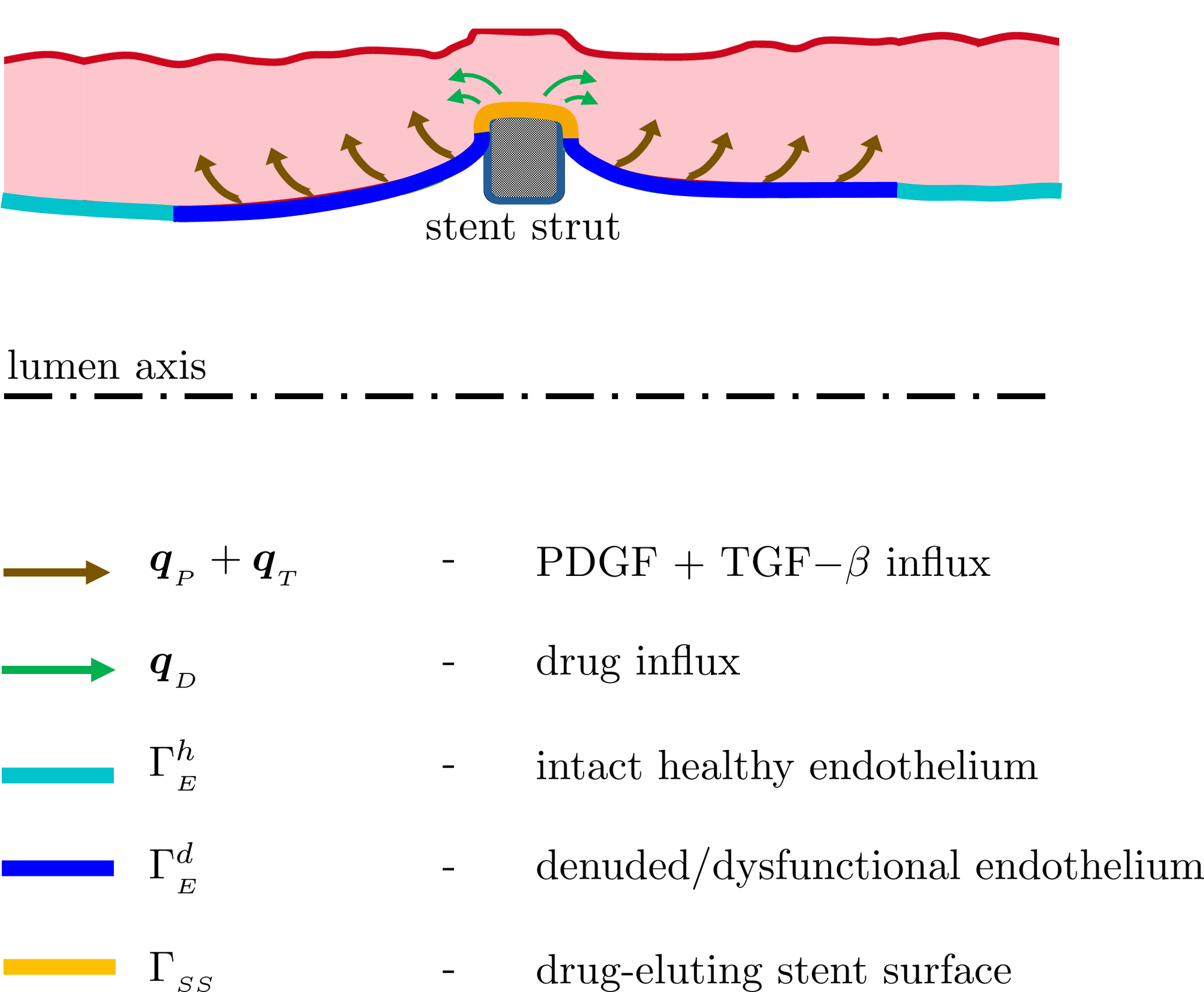}
    \caption{\textbf{Boundary conditions}}
    \label{fig_BC}
\end{figure}

\begin{table}[htbp!]\
\centering
\caption{\textbf{Boundary conditions}}
\label{table_bc} 
\begin{tabular}{p{0.2cm}p{1.2cm}p{1.2cm}p{3cm}p{6cm}p{4cm}}
\noalign{\hrule height 0.05cm}\noalign{\smallskip}
& variable & Dirichlet & region & Neumann & region \\
\noalign{\hrule height 0.05cm}\noalign{\smallskip}
\ldelim\{{7}{0.2mm}[(a)] &\,\,   $c_{{}_P}$ & $-$ & $-$ & \(\bm{q}_{{}_P}\cdot\bm{n} = \bar{q}_{{}_P} :=  f_{{}_{B1}}(t)\,\displaystyle{\left(1 - \frac{\rho_{{}_{E}}}{\rho_{{}_{E,eq}}}\right)}\,\bar{q}^{\text{ref}}_{{}_{P}}\) & $\forall \bm{x}\in \Gamma_{{}_E} $ \\
\\
&  &  & & $\text{\sf grad}\,(c_{{}_P})\cdot\bm{n} = 0$ &  \( \forall \bm{x}\in  \Gamma \,\symbol{92}\, \Gamma_{{}_E} \)\\
\\
&\,\,   $c_{{}_T}$ & $-$ & $-$ & \(\bm{q}_{{}_T}\cdot\bm{n} = \bar{q}_{{}_T} :=  f_{{}_{B1}}(t)\,\displaystyle{\left(1 - \frac{\rho_{{}_{E}}}{\rho_{{}_{E,eq}}}\right)}\,\bar{q}^{\text{ref}}_{{}_{T}}\) & $\forall \bm{x}\in \Gamma_{{}_E} $\\
\\
&  &  & & $\text{\sf grad}\,(c_{{}_T})\cdot\bm{n} = 0$ & \( \forall \bm{x}\in  \Gamma \,\symbol{92}\, \Gamma_{{}_E} \)\\
\\
\ldelim\{{5}{0.2mm}[(b)]&\,\,   $c_{{}_D}$ & $-$ & $-$ & \(\bm{q}_{{}_{D1}}\cdot\bm{n} = \bar{q}_{{}_{D1}} :=  f_{{}_{B2}}(t)\,\bar{q}^{\text{ref}}_{{}_{D}}\) & \(\forall \bm{x}\in\Gamma_{{}_{SS}}\) \\
\\
& & & & \(\bm{q}_{{}_{D2}}\cdot\bm{n} = \bar{q}_{{}_{D2}} :=  - \varepsilon_{{}_{D2}}\,\rho_{{}_E}\,c_{{}_D}\) & \(\forall \bm{x}\in\Gamma_{{}_E} \) \\
\\
&  &  & & $\text{\sf grad}\,(c_{{}_D})\cdot\bm{n} = 0$ & $\forall \bm{x}\in \Gamma \,\symbol{92}\, \left(\Gamma_{{}_{SS}}\cup\Gamma_{{}_E} \right)$\\
\\
\ldelim\{{5}{0.2mm}[(c)]& \,\, $c_{{}_C}$ & $-$ & $-$ & \(\text{\sf grad}\,(c_{{}_C})\cdot\bm{n} = 0\) & \(\forall \bm{x} \in \Gamma\)\\
\\
& \,\, $\rho_{{}_S}$ & $-$ & $-$ & \(\text{\sf grad}\,(\rho_{{}_S})\cdot\bm{n} = 0\) & \(\forall \bm{x} \in \Gamma\)\\
\\
&\,\, $\rho_{{}_{E}}$ & $-$ & & \(\text{\sf grad}\,(\rho_{{}_{E}})\cdot\bm{n} = 0 \) & \( \forall \bm{x}\in\partial \Gamma_{{}_{E}}\)\\
\\
\,\,(d) &\,\, $\bm{u}$ & $\bm{u} = \bar{\bm{u}}$ & $\forall \bm{x}\in \Gamma_{0,u}$ & \(\bm{T} = \bm{P}\cdot\bm{N} = \bar{\bm{T}}\) & \(\forall \bm{x}\in \Gamma_{0,T}\)\\
\noalign{\hrule height 0.05cm}\noalign{\smallskip}
\end{tabular}
\end{table}

All the relevant boundary conditions are summarized in Table \ref{table_bc}. The whole boundary of the arterial wall $\Gamma$ is first divided into the boundary exposed to the lumen $\Gamma_{{}_l}$, and the remaining boundary $\Gamma_{{}_r}$, i.e., $\Gamma = \Gamma_{{}_l} \cup \Gamma_{{}_r}$. Further, $\Gamma_{{}_l}$ is partitioned into the endothelium $\Gamma_{{}_{E}}$ and drug-eluting stent surface $\Gamma_{{}_{SS}}$, i.e., $\Gamma_{{}_l} = \Gamma_{{}_{E}} \cup \Gamma_{{}_{SS}}$. Finally, the endothelium $\Gamma_{{}_{E}}$ is subdivided into the transient boundaries which represent the intact healthy endothelium ($\Gamma^h_{{}_{E}}$) and the denuded/dysfunctional endothelium ($\Gamma^d_{{}_{E}}$) respectively, i.e.,  $\Gamma_{{}_{E}} = \Gamma^h_{{}_{E}} \cup \Gamma^d_{{}_{E}}$. The boundaries lining the endothelium are just denoted $\partial \Gamma_{{}_{E}}$.

\begin{enumerate}
\item[(a)] Through the process of platelet aggregation, activation, and degranulation, PDGF and TGF-$\beta$ enter the arterial wall at sites of  endothelial damage exacted by stent implantation as well as at sites of dysfunctional endothelial barrier function induced by the blood flow dynamics and the pharmacodynamics ($\Gamma_{{}_E}$  in Fig. \ref{fig_BC}). PDGF and TGF-$\beta$ hence shall have their respective influxes prescribed in the form
\begin{eqnarray}
    \bm{q}_{{}_P}\cdot\bm{n} &=& \bar{q}_{{}_P}  := f_{{}_{B1}}(t)\,\displaystyle{\left(1 - \frac{\rho_{{}_E}}{\rho_{{}_{E,eq}}}\right)}\,\bar{q}^{\sf ref}_{{}_P} \nonumber\\
    \bm{q}_{{}_T}\cdot\bm{n} &=& \bar{q}_{{}_T} := f_{{}_{B1}}(t)\,\displaystyle{\left(1 - \frac{\rho_{{}_E}}{\rho_{{}_{E,eq}}}\right)}\,\bar{q}^{\sf ref}_{{}_T}\,\, \quad \forall \bm{x}\in \Gamma_{{}_E} 
\end{eqnarray}
where $\bar{q}^{\text{ref}}_{{}_{P}}$ and $\bar{q}^{\text{ref}}_{{}_{T}}$ are parameters controlling the peaks of the respective influx profiles. $f_{{}_{B1}}$ is a time ($t$)-dependent scaling function defined as
\begin{equation}
        f_{{}_{B1}}(t) := 1 - \exp(-l_{{}_{B}}\,t)
\end{equation}
where the parameter $l_{{}_{B}}$ modulates the slope of the influx profiles, thereby indirectly prescribing the time required to reach the peak influxes $\bar{q}^{\text{ref}}_{{}_{P}}$ and $\bar{q}^{\text{ref}}_{{}_{T}}$, and $f_{{}_{B1}}(t)$ behaves qualitatively similar to $f_{{}_{E2}}(c_{{}_D})$ (see Fig. \ref{fig_fE2_cD}). The dependence of both $\bar{q}_{{}_P}$ and $\bar{q}_{{}_T}$ on the EC density ($\rho_{{}_E}$) is prescribed so as to render the influx of growth factors dependent on the health of the endothelium. The larger the deviation from the homeostatic value of EC density ($\rho_{{}_{E,eq}}$), the more the dysfunctional behavior of the endothelium, allowing for more platelet aggregation and influx of PDGF and TGF-$\beta$, and enhanced monocyte infiltration.
Homogeneous Neumann boundary conditions are applied on the remaining boundaries.
\item[(b)] Drug from the DES is assumed to directly diffuse into the vessel wall regions of stent apposition against the vessel wall. Hence an influx of the drug is prescribed on the aforementioned contact surface ($\Gamma_{{}_{SS}}$ in Fig. \ref{fig_BC}) in the form
\begin{equation}
    \bm{q}_{{}_{D1}}\cdot\bm{n} = \bar{q}_{{}_{D1}} := f_{{}_{B2}}(t)\,\bar{q}^{\text{ref}}_{{}_{D}}, 
\end{equation}
where $q^{\text{ref}}_{{}_D}$ is a parameter controlling the peak of the influx profile, while $f_{{}_{B2}}$ is a time ($t$)-dependent factor defined to be
\begin{eqnarray}\label{load-factor_drug}
f_{{}_{B2}}(t) &:=& \displaystyle{\exp \left(-\frac{t}{f_{{}_{B3}}(t_{{}_c})}\right)\,\left(1 - \exp \left(-\frac{t}{t_{{}_c}}\right)\right)},\\
&\text{wherein}&\nonumber\\
    f_{{}_{B3}}(t_{{}_c}) &:=& \displaystyle{t_2\,\exp\left(\frac{t_p}{t_{{}_c}}\right) - t_{{}_c}}.
\end{eqnarray}
Here, $t_{{}_p}$ and $t_{{}_c}$ are time-like parameters that control the curvature of the influx profile, and $t_p$ can be interpreted to be the time to achieve peak influx of the drug. Exemplary drug influx profiles and the associated cumulative drug release profiles are shown in Fig. \ref{fig_drp}.
\begin{figure}[!htb]
    \centering
    \begin{minipage}{.5\textwidth}
        \centering
        \includegraphics[scale = 0.6]{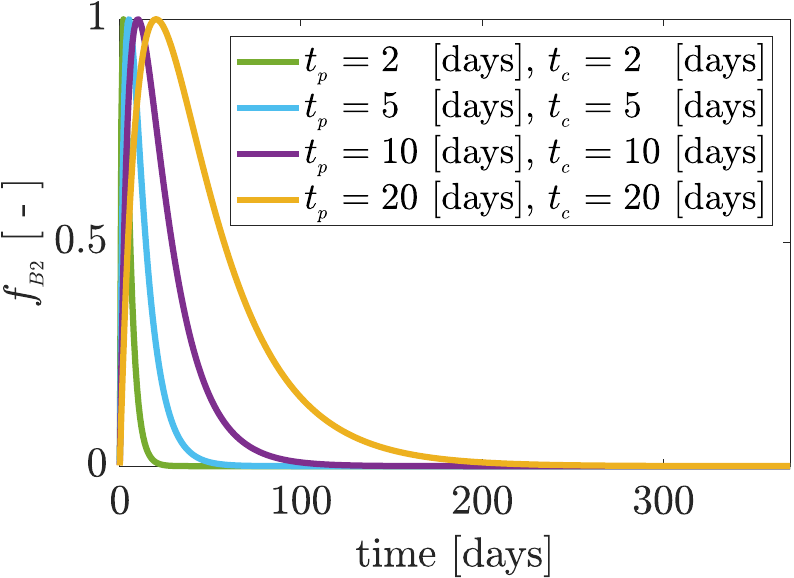}
    \end{minipage}%
    \begin{minipage}{0.5\textwidth}
        \centering
        \includegraphics[scale = 0.6]{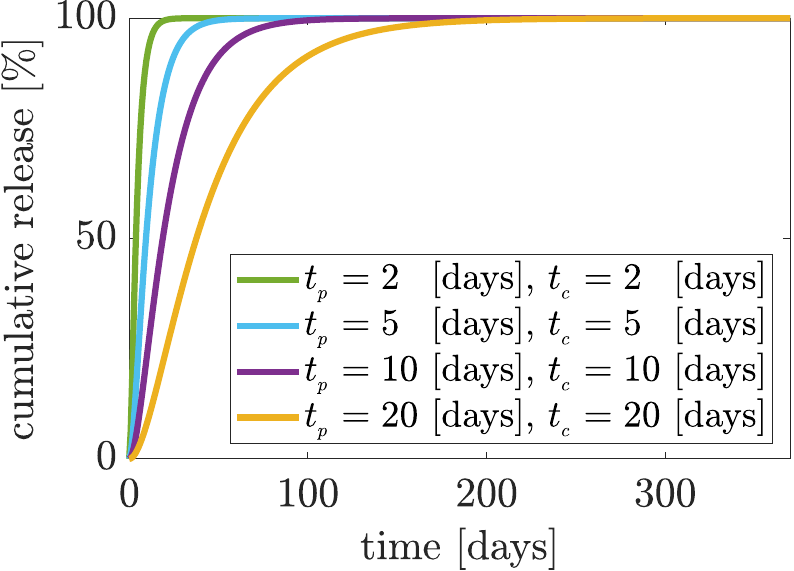}
    \end{minipage}
    \caption{\textbf{ left - drug influx profiles; right - cumulative release profiles} }
    \label{fig_drp}
\end{figure}
\vspace{0.1in}\\
\noindent The binding of the drug to the FKBP12 receptors on the ECs is modeled via a flux boundary condition on the luminal surface of the form
\begin{equation}
    \bm{q}_{{}_{D2}}\cdot\bm{n} = \bar{q}_{{}_{D2}} := - \varepsilon_{{}_{D2}}\,\rho_{{}_{E}}\,c_{{}_D},
\end{equation}
$\varepsilon_{{}_{D2}}$ being termed the EC-pertinent receptor internalization coefficient for the drug. Homogeneous Neumann boundary conditions are applied on the remaining boundaries.

\item [(c)] ECM ($c_{{}_E}$), SMCs ($\rho_{{}_S}$) and ECs ($\rho_{{}_{en}}$) are all considered constrained within the arterial wall. Therefore homogeneous Neumann boundary conditions are prescribed for the respective balance equations on the entire boundary of the vessel wall.

\item[(d)] Displacements are prescribed on the boundary $\Gamma_{{0,u}}$ in the reference configuration, and tractions on the boundary $\Gamma_{{0,T}}$ in the reference configuration. Also, the total boundary in the reference configuration $\Gamma_{0} = \Gamma_{{0,u}} \cup \Gamma_{{0,T}}$.
\end{enumerate} 

\subsection{Initial conditions}\label{ic}

The initial ECM concentration ($c_{{}_C}$) and SMC density ($\rho_{{}_S}$) are prescribed to be those of a healthy homeostatic artery in equilibrium in the entire domain of the arterial wall. PDGF, TGF$-\beta$, and the drug are considered initially absent in the vessel wall. EC density ($\rho_{{}_{E}}$) is prescribed to be zero at regions where endothelial denudation is observed/expected ($\Gamma^d_{{}_{E}}$), and to be that on a healthy homeostatic artery in equilibrium ($\rho_{{}_{E,eq}}$) elsewhere ($\Gamma^h_{{}_{E}}$). Table \ref{table_ic} summarizes the relevant initial conditions.

\begin{table}[htbp!]\
\centering
\caption{\textbf{Initial conditions}}
\label{table_ic} 
\begin{tabular}{p{2.cm}p{2.5cm}p{2cm}}
\noalign{\hrule height 0.05cm}\noalign{\smallskip}
variable &  initial condition & region\\
\noalign{\hrule height 0.05cm}\noalign{\smallskip}
$c_{{}_P}$ & $0$ & \(\forall \bm{x} \in \Omega\)\\
\\
$c_{{}_T}$ & $0$ & \( \forall \bm{x} \in \Omega\) \\
\\
$c_{{}_C}$ & $c_{{}_{C}}$ & \( \forall \bm{x} \in \Omega\)\\
\\
$c_{{}_D}$ & $0$ & \( \forall \bm{x} \in \Omega\)\\
\\
$\rho_{{}_S}$ & $\rho_{{}_{S,eq}}$ & \( \forall \bm{x} \in \Omega\)\\
\\
$\rho_{{}_{E}}$ & 0 & \( \forall \bm{x} \in \Gamma^d_{{}_{E}}\)\\
\\
$\rho_{{}_{E}}$ & $\rho_{{}_{E,eq}}$ & \( \forall \bm{x} \in \Gamma^h_{{}_{E}}\)\\
\\
\noalign{\hrule height 0.05cm}\noalign{\smallskip}
\end{tabular}
\end{table}
\vspace{-0.1in}

\section{Finite element implementation}\label{fe_impl}
The coupled differential equations, i.e., Eqs. \ref{pdgf_bal}, \ref{tgf_bal}, \ref{ce_bal}, \ref{drug_bal}, and \ref{smc_bal}, are first transformed to their Lagrangian equivalents as done by \citet{MANJUNATHA2022106166}. All the balance equations, except that of the EC density ($\rho_{{}_{E}}$), in their weak forms, are solved in the arterial wall domain, spatially discretized by trilinear hexahedral elements. The EC density ($\rho_{{}_{E}}$) field is solved for on the lumen surface, spatially  discretized using bilinear quadrilaterals projected from the bulk arterial wall mesh. This projected surface mesh is also utilized in prescribing flux (inhomogeneous Neumann) boundary conditions. Temporal discretization is performed utilizing the fully-implicit backward-Euler method. Additionally, for the SMC density ($\rho_{{}_S}$) field, the streamline-upwind Petrov-Galerkin (SUPG) method is incorporated in order to ensure robust finite element computations at high chemotactic and haptotactic velocities of the SMCs. The system matrices intertwine with each other at the common nodes shared between the bulk hexahedrons and the surface quadrilaterals. The framework presented is implemented into the commercial finite element program FEAP via its user-defined element interface. The tangential stiffness matrices involved in each global Newton iteration of the nonlinear problem are obtained via algorithmic differentiation using the \textit{Wolfram Mathematica} package \textit{AceGen} \cite{korelc2002,korelc2009}. The solution of the linear system of equations in every global Newton iteration is efficiently accomplished with the help of the parallelized  PARDISO solver \cite{pardiso-7.2a,pardiso-7.2b,pardiso-7.2c} through a user-defined macro.

%The resulting system of linear equations as part of the monolithic solution construct shall entail a tangential stiffness matrix at the element level of the form
%\begin{equation}\label{stiff_mat_bulk}
%\begin{bmatrix}
    %\bm{K}^e_{{}_{PP}} & \bm{K}^e_{{}_{PT}} & \bm{0} & \bm{K}^e_{{}_{PD}} & \bm{K}^e_{{}_{PS}} &\bm{K}^e_{{}_{Pu}}\\
    %\bm{0} & \bm{K}^e_{{}_{TT}} & \bm{0} & \bm{0} & \bm{K}^e_{{}_{TS}} & \bm{K}^e_{{}_{Tu}} \\
    %\bm{K}^e_{{}_{CP}} & \bm{0} & \bm{K}^e_{{}_{CC}} & \bm{0}  & \bm{K}^e_{{}_{CS}} & \bm{K}^e_{{}_{Cu}} \\
    %\bm{0} & \bm{K}^e_{{}_{DD}} & \bm{0} & \bm{0} & \bm{K}^e_{{}_{DS}} &  \bm{K}^e_{{}_{Du}} \\
    %\bm{K}^e_{{}_{SP}} & \bm{K}^e_{{}_{ST}} & \bm{K}^e_{{}_{SC}} & \bm{K}^e_{{}_{SD}} & \bm{K}^e_{{}_{SS}} & \bm{K}^e_{{}_{Su}} \\
    %\bm{0} & \bm{0} & \bm{K}^e_{{}_{uC}} & \bm{0} & \bm{K}^e_{{}_{uS}} & \bm{K}^e_{{}_{uu}}
%\end{bmatrix}
%\end{equation}
%for the bulk hexahedral elements, while it is of the form
%\begin{equation}\label{stiff_mat_flux_int}
%\begin{bmatrix}
    %\bm{K}^e_{{}_{P\,\,en}} & \bm{0} & \bm{K}^e_{{}_{Pu}} \\
    %\bm{K}^e_{{}_{T\,\,en}} & \bm{0} & \bm{K}^e_{{}_{Tu}} \\
    %\bm{K}^e_{{}_{D\,\,en}} & \bm{K}^e_{{}_{DD}} & \bm{K}^e_{{}_{Pu}} \\
    %\bm{K}^e_{{}_{en\,\,en}} & \bm{K}^e_{{}_{en\,\,D}} & %\bm{K}^e_{{}_{en\,\,u}}
%\end{bmatrix}
%\end{equation}
%for the projected quadrilaterals on the luminal side of the arterial wall $\Gamma_{{}_l}$. 

\section{Numerical evaluation}\label{sect_num_eval}
To evaluate the computational model presented herein, two boundary value problems (BVPs) are set up. The first BVP presents a simplified representation of a stented arterial quadrant with partial endothelial denudation, utilizing which the spatiotemporal evolution of the mediators of restenosis are examined. Additionally, the respective influences of the severity in the inflammatory response post stenting as well as the level of drug embedment in the stent struts are evaluated. Following this, a longitudinal section of a coronary artery implanted with the XIENCE-V stent is considered for the numerical evaluation. In both the BVPs, media and adventitia are considered to constitute the bulk of the arterial wall, while the endothelium is modeled as a surface on the luminal side. Arterial overstretch is ignored in this context since the denudation of the endothelium is considered the key driver for the restenotic process.

\subsection{Simplified stented arterial segment}\label{ssas}

\begin{figure}[htbp]
    \centering
    \includegraphics[scale=0.6]{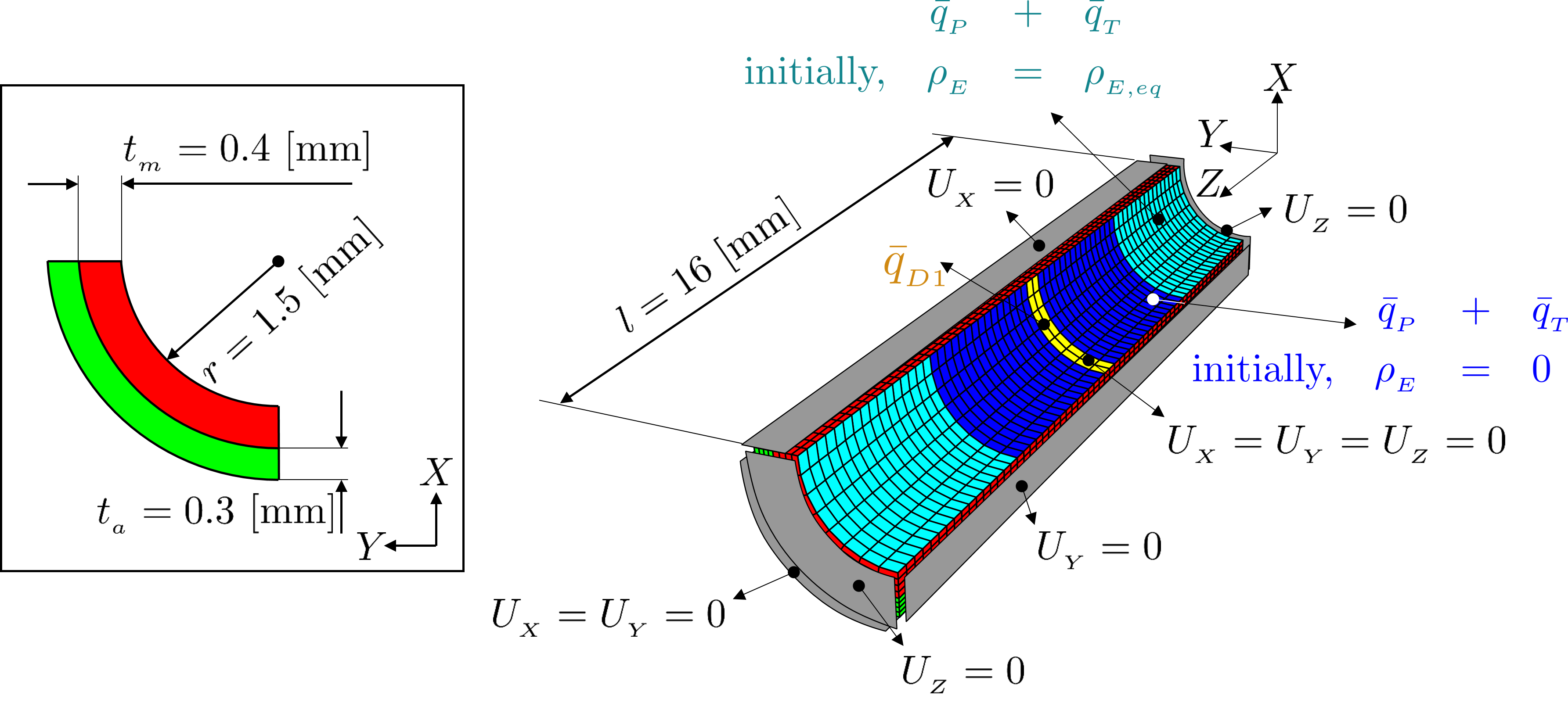}
    \caption{\textbf{Simplified stented arterial segment}}
    \label{stented_segment_simp}
\end{figure}

A quadrant of an artery of length $l = 16$ [mm], lumen radius $r = 1.5$ [mm], medial thickness $t_{{}_m} = 0.4$ [mm], and adventitial thickness $t_{{}_a} = 0.3$ [mm] is constructed as shown in Fig. \ref{stented_segment_simp}. A segment of length 0.5 [mm] on the luminal surface of the arterial wall at mid-length ($Z = 8$ [mm]) is considered to be the region of stent apposition ($\Gamma_{{}_{SS}}$). Regions of length 3.5 [mm] on each side of the stent apposition are considered denuded of endothelial cells ($\Gamma^d_{{}_{E}}$). The rest of the inner surface of the vessel wall is considered to possess a healthy endothelium ($\Gamma^h_{{}_{E}}$).

\subsubsection{Discretization}
The geometry is spatially discretized with trilinear hexahedral elements, 64 of them along the longitudinal direction, 8 along the circumference, and 4 along the radial direction of each layer of the arterial wall. Projected surface meshes on the luminal side represent the stent apposition region as well as the denuded and healthy endothelial regions. 
The problem is temporally discretized with a time-step $\Delta t$ of 1 [days].

\subsubsection{Boundary and initial conditions}
Displacement boundary conditions mimicking the symmetry of the arterial wall are prescribed. The ends are fixed against longitudinal movement. The abluminal side of the vessel wall is constrained against displacements in the circumferential and radial directions. The boundary and initial conditions are prescribed as elucidated in Section \ref{bc} on respectively annotated regions in Fig. \ref{stented_segment_simp}.

%Drug influx $\bar{q}_{{}_D}$ is prescribed onto the stent apposition region $\Gamma_{{}_{SS}}$. PDGF influx $\bar{q}_{{}_P}$ and TGF-$\beta$ influx $\bar{q}_{{}_T}$ are prescribed onto regions of denuded endothelium $\Gamma^d_{{}_{en}}$. The surfaces $\Gamma^d_{{}_{en}}$ and $\Gamma^h_{{}_{en}}$ vary spatially based on the temporal evolution of the EC density ($\rho_{{}_{en}}$) field.  

%EC density $\rho_{{}_{en}}$ is prescribed initially to be zero on  $\Gamma^d_{{}_{en}}$, and to be that on a healthy homeostatic artery in equilibrium $\rho_{{}_{en,eq}}$ on $\Gamma^h_{{}_{en}}$.

\subsubsection{Parameters}
Out of the 39 parameters listed in Tables \ref{model_params_pdgf} to \ref{model_params_EC}, 20 are carried over from relevant literature. The remaining 19 are chosen in such a way that the transient response of the mediators and the macroscopic outcome is physiological.

\subsubsection{Results and discussion}
Three aspects of pharmacokinetics and pharmacodynamics are evaluated using the simplified model setup. 
\begin{enumerate}
    \item The level of drug embedment in the stent strut, controlled via the drug influx parameter $\bar{q}^{\text{ref}}_{{}_D}$

    \item Drug release rate prescribed by the release profiles depicted in Fig. \ref{fig_drp}

    \item Severity of patient-specific inflammatory response, controlled by the PDGF secretion ratio $r_{{}_{\eta}}$
\end{enumerate}

\begin{figure}[htbp!]
    \centering
    \includegraphics[scale=0.75]{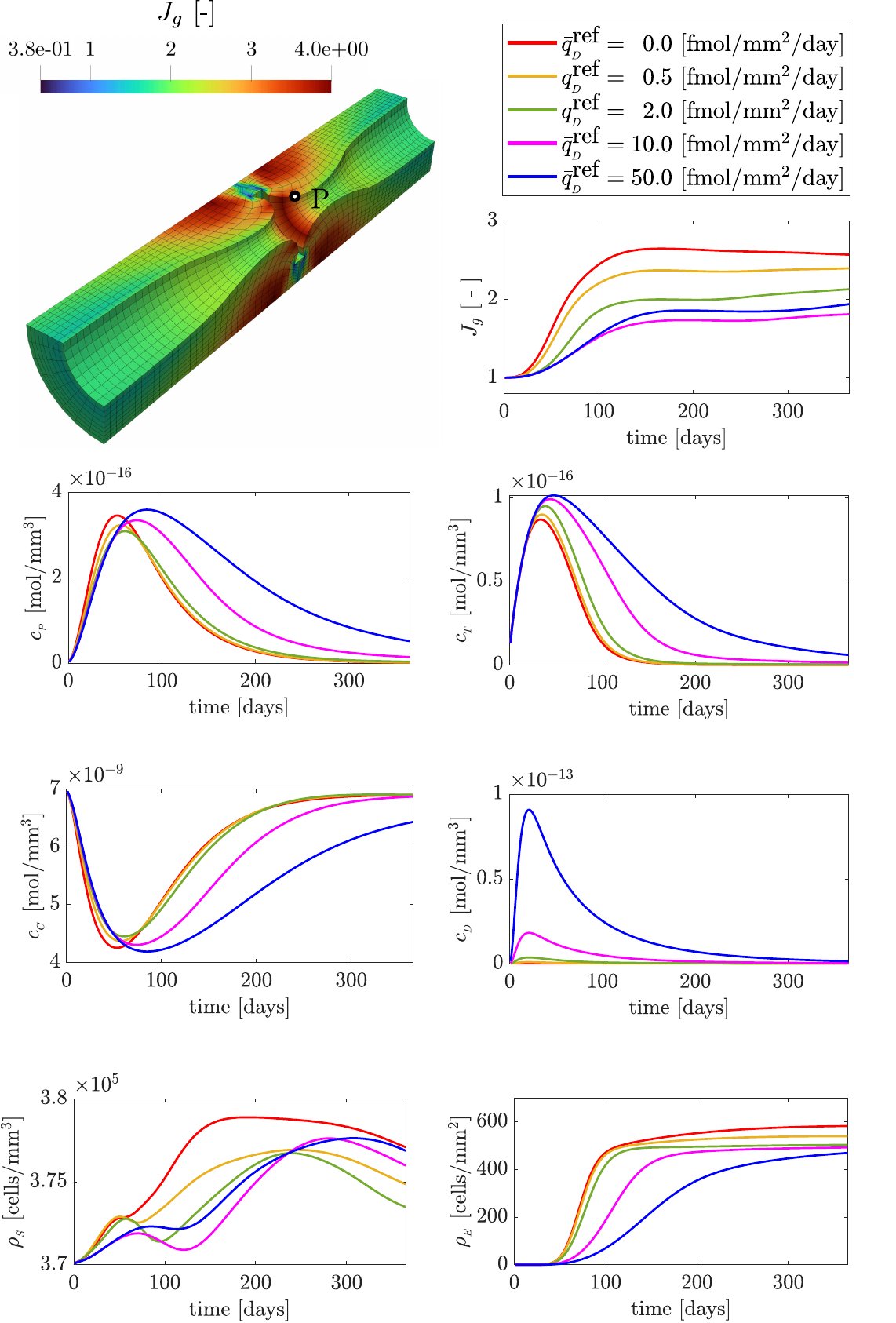}
    \caption{\textbf{Evolution of quantities of interest at point P}}
    \label{evolutions}
\end{figure}

\myparagraph{Influence of the level of drug embedment in the stent strut}
Fig. \ref{evolutions} portrays how at point P the evolution of the growth stretch $\vartheta = J_g$ and the mediators of the restenotic growth is influenced by $\bar{q}^{\text{ref}}_{{}_D}$ for a fixed release profile and a PDGF secretion ratio of $r_{{}_{\eta}} = 0.5$. It is interesting to note that a nonlinear behavior is observed in the growth evolution for increasing levels of drug embedment. For this particular BVP and the choice of parameters, $\bar{q}^{\text{ref}}_{{}_D} = 10 $ [fmol/mm$^2$/day] is determined to be the optimal peak drug influx across the stent contact surface. Beyond this, the apoptotic and antiproliferative mechanisms induced in the ECs due to the drug become predominant, thereby slowing down the healing of the endothelium and allowing for more platelet degranulation and monocyte infiltration into the subendothelial space. The delay in endothelial healing is clearly observable in the evolution of $\rho_{{}_{E}}$. This result is in contradiction to the investigation presented by \citet{MCQUEEN2022992}, wherein the restenotic growth was observed to be monotonically decreasing and reaching an asymptotic limit beyond which there was no observable difference in thickness of the neointimal tissue. Although the authors therein consider a highly resolved drug transport and SMC proliferation model, due to the neglection of pharmacodynamic effects on ECs in their model, this disparity is justified.

\begin{figure}[htbp]
    \centering
    \includegraphics[scale=1.2]{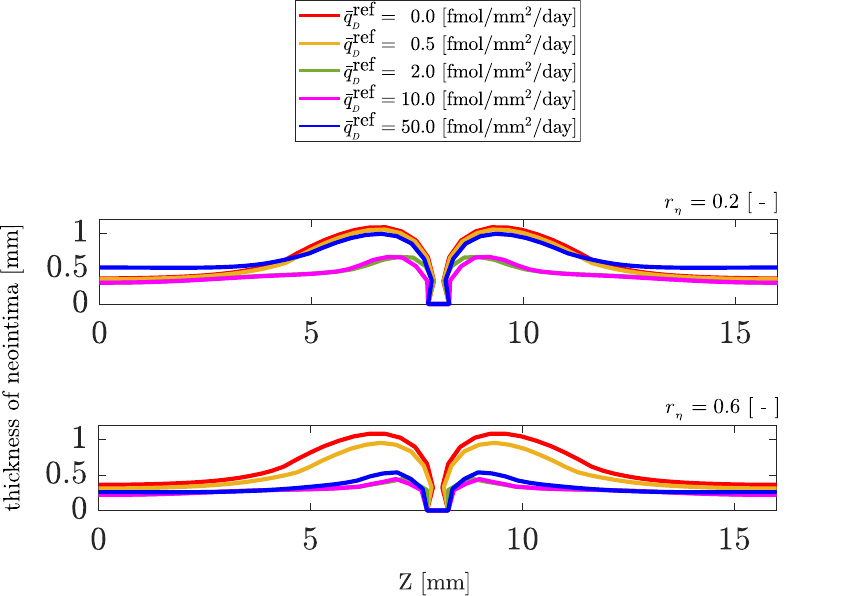}
    \caption{\textbf{Neointima profiles for varying PDGF secretion ratios }}
    \label{neointima_comp_reta}
\end{figure}

\myparagraph{Influence of the drug release profile}
Restenotic growth is investigated for two levels of drug embedment and varying drug release profiles (see Fig. \ref{fig_drp}), and the resulting neointima profiles are plotted in Fig. \ref{neointima_comp_reta}. At low levels of drug mass embedded in the stent strut, a slow release of the drug into the arterial wall results in the least neointimal growth. But at the optimal drug embedment associated with $\bar{q}^{\text{ref}}_{{}_D} = 10 $ [fmol/mm$^2$/day], the influence of the release profiles is negligible. However, a non-monotonic behavior is observed, as can be seen in the magnified images. This non-monotonicity is also what \citet{MCQUEEN2022992} discerned in their investigations but the trend was consistent across all drug masses in their work, whereas a shift is observed herein.

\begin{figure}[htbp]
    \centering
    \includegraphics[scale=1.2]{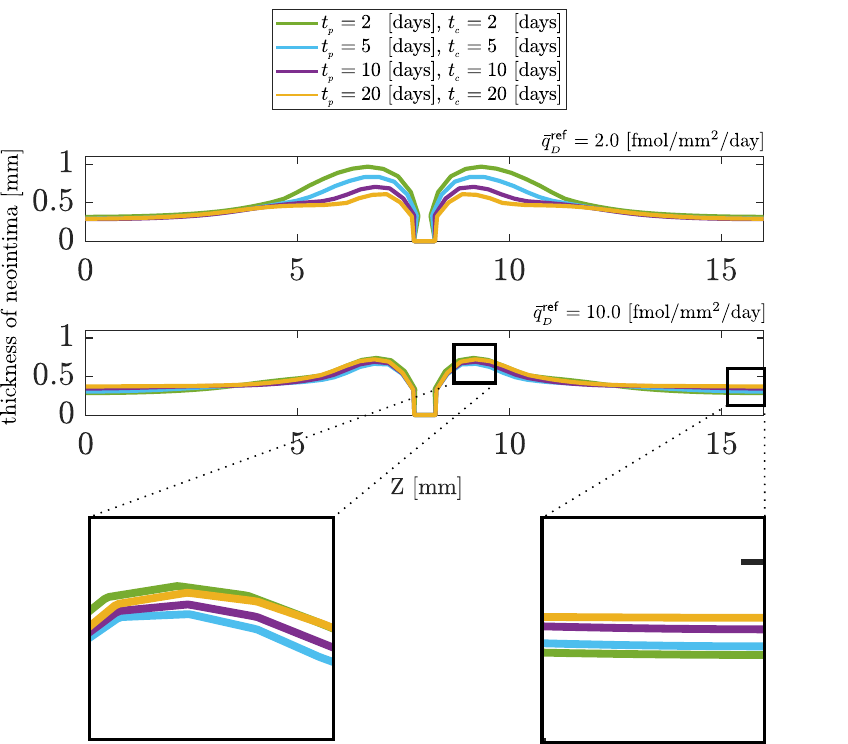}
    \caption{\textbf{Neointima profiles for varying drug influx profiles (see Fig. \ref{fig_drp})}}
    \label{neointima_comp_rp}
\end{figure}

\myparagraph{Influence of severity of the inflammatory response}
The inflammatory response, specifically the expression of cell adhesion molecules (ICAM-1 and VCAM-1) on the endothelial surface, highly influences the pharmacodynamics \citep{DANIEL201779} since they alter the amount of infiltrating monocytes, and consequently, the amount of PDGF sourced within the vessel wall. This effect is clearly seen in Fig. \ref{neointima_comp_reta} where the neointima profiles for two different values of $r_{{}_{\eta}}$ are presented for varying drug embedment levels. The optimal range of the embedded drug mass becomes wider and more effective at high values of $r_{{}_{\eta}}$ since the anti-inflammatory effects of the drug become highly influential. In all cases, an increased drug presence beyond an optimal range shall drastically enhance the inflammation and thereby the restenotic growth, almost equivalent to levels observed after BMS implantation.

\subsection{Coronary artery implanted with XIENCE-V stent}
\begin{figure}[htbp!]
    \centering
    \includegraphics[scale=0.75]{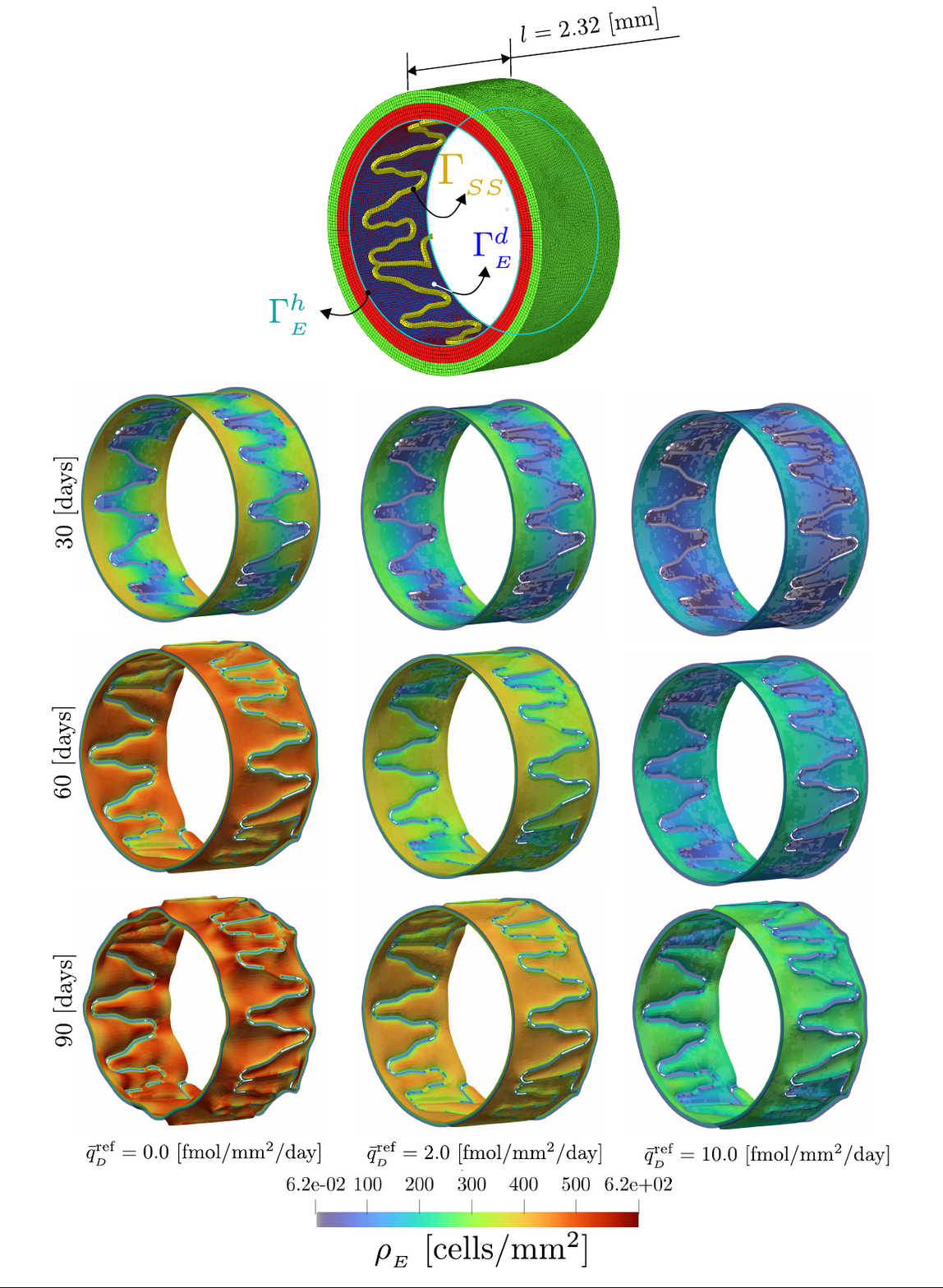}
    \caption{\textbf{Evolution of endothelial health}}
    \label{fig_stented_artery}
\end{figure}

A segment of the human coronary artery of length $l =2.32$ [mm], implanted with the XIENCE-V stent, is constructed using the stent geometry as shown in the top in Fig. \ref{fig_stented_artery}. This segment is chosen since it repeats itself longitudinally throughout the length of the stent. 

\subsubsection{Discretization}
The geometry of the arterial wall is spatially discretized with 99,368 trilinear hexahedral elements in the bulk, 4 elements along the thickness of each layer. The stent apposition surface, projected from the bulk, contains 2054 bilinear quadrilateral elements. The rest of the luminal surface then consists of 10,367 bilinear quadrilaterals. Temporal discretization is again in terms of $\Delta t = 1$ [days].

\subsubsection{Boundary and initial conditions}
The stented surface is constrained against all displacements, while the exterior surface of the geometry is constrained against movements in circumferential and radial directions as is the case in Section \ref{ssas}. The color coding for the discretized representation at the top of Fig. \ref{fig_stented_artery} is maintained to be the same as in Fig. \ref{stented_segment_simp} to emphasize the fact that all flux boundary conditions are applied exactly as in Section \ref{ssas} on the corresponding boundary domains. The only difference lies in the prescription of the initial values of EC density ($\rho_{{}_{E}}$). While the entire luminal surface is considered denuded, to initiate the healing of the endothelium, the values for $\rho_{{}_{E}}$ are prescribed initially to be $\rho_{{}_{E,eq}}$ along the annular edges at the longitudinal ends of the surface, and it is set to be zero elsewhere.

\subsubsection{Parameters}
The model parameters remain the same as in the previous numerical example.

\subsubsection{Results and discussion}
A period of 90 [days] after stent implantation was simulated using the BVP setup. This is due to the fact that stent strut coverage by neointimal is not modeled as part of the presented methodology. In this regard, the obstruction of the stent apposition against restenotic growth is artificial. The results hence serve only as a reflection of the capability of the model to capture the intricate pathophysiology within complex geometric settings.

\begin{figure}[htbp!]
    \centering
    \includegraphics[scale=0.75]{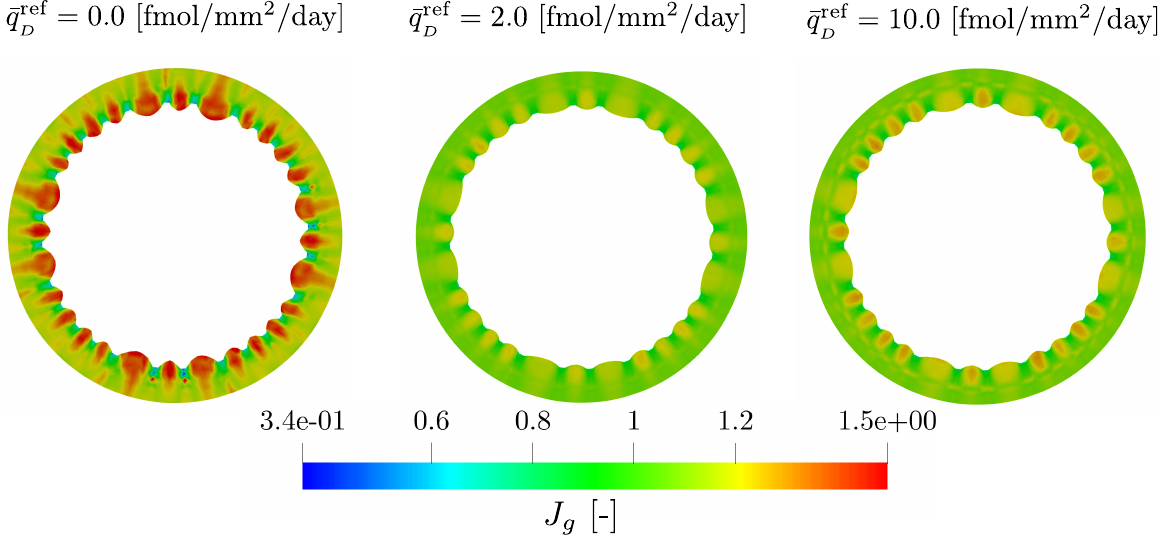}
    \caption{\textbf{Mid-length slices of growth stretch at 90 [days]}}
    \label{fig_stented_artery_slice}
\end{figure}

The collage in Fig. \ref{fig_stented_artery} presents the transient process of endothelial healing through 90 [days] for different levels of drug embedment. As expected, the restenotic growth is much pronounced in the case of bare metal stent struts, while also the endothelial recovery is relatively faster compared to DES. Additionally, drug mass optimality is observed at $\bar{q}^{\text{ref}}_{{}_D} = 2.0 $ [fmol/mm$^2$/day] beyond which there is an increase in neointimal growth. This is clearly demonstrated in Fig. \ref{fig_stented_artery_slice} where mid-length slices (at $l = 1.16$ [mm]) representing the growth stretch $J_g$ are plotted.

\section{Conclusion and outlook}\label{sect_conclusion}
The current work presents a computational modeling framework to model the spatiotemporal evolution of neointimal growth observed during the pathophysiology of in-stent restenosis. The main contribution of the work lies in capturing the regulatory effects of the endothelium in stented sections of arteries, endothelial recovery, and the pharmacological influence of rapamycin-based drugs on the very intricate restenotic process. The transient evolutions of the significant mediators, including the ECs and the drug, are then coupled to a transversely isotropic constitutive description of the arterial wall through  thermodynamically consistent growth kinematics. The framework can model the complex interplay between the mediators, providing significant resolution into cellular and biochemical interactions. Through the numerical investigations, it has been demonstrated that there is scope for optimization of the drug content embedded within the struts of DESs as well as the drug release profiles adopted. Finally, the influence of the patient-specific immunological response that governs the severity of the inflammatory response is observed to bear a significant influence on the aforementioned optimization.

It is to be noted that the model presented considers the denudation of endothelium to be the only driver of restenotic growth. While it does play a significant role, it is not the only source of neointimal hyperplasia. The arterial overstretch is diagnostically determined to be the other important aspect that has to be considered in the modeling approach. In this context, implementing a damage-dependent multiphysical model developed by \citet{GIERIG2023106811} shall render a significantly enhanced resolution to the computational model. The other crucial aspect that has to be dealt with is the effect of blood dynamics, particularly the arterial wall shear stress, and pharmacokinetics of the drug that leads to downstream deposition owing to the lipophilic nature of the drug. Embedding the hence fully resolved multiphysics-based model within a fluid-structure interaction (FSI) framework shall then enable accurate patient-specific risk prediction and pharmacological evaluation.

The predicament lies in validating the complex computational model through suitable experiments. Although several of the modeling parameters are extracted from literature which presents immunohistochemical assays, the rest of them have been chosen based on physiological insights and expected outcomes. The computational model hence serves as a tool to pick possible avenues for experimental validation, the identification of which is based currently either on physiological intuition or extensive data analytics. 
\appendix
\appendixpage
\addappheadtotoc
\section{Clausius-Duhem inequality}\label{cde}

Using Eqs. \ref{psidot}, \ref{cedot}, \ref{cgdot}, and \ref{hbardot} in the Clausius-Duhem inequality (Eq. \ref{cde1}), we arrive at the reduced dissipation inequality in the form
\begin{eqnarray}
    \mathscr{D}_{{}_{\sf red}} &:=&\left(2\,\bar{\bm{C}}_e\,\frac{\partial \bar{\psi}}{\partial \bar{\bm{C}}_e} - 2\,\frac{\partial \bar{\psi}}{\partial \bm{C}_g}\,\bm{C}_g - 2\,\frac{\partial \bar{\psi}}{\partial \bar{\bm{H}}_1}\,\bar{\bm{H}}_1 - 2\,\frac{\partial \bar{\psi}}{\partial \bar{\bm{H}}_2}\,\bar{\bm{H}}_2 + c^{0}_{{}_C}\,\frac{\partial \bar{\psi}}{\partial c^0_{{}_C}}\,\bm{I}\right):\bar{\bm{L}}_g\nonumber\\
    \nonumber\\
    & & - \left(\bm{U}_g\,\frac{\partial \bar{\psi}}{\partial \bar{\bm{H}}_1}\,\bm{U}_g\right):\dot{\bm{H}}_1 - \left(\bm{U}_g\,\frac{\partial \bar{\psi}}{\partial \bar{\bm{H}}_2}\,\bm{U}_g\right):\dot{\bm{H}}_2\nonumber\\
    \nonumber\\
    & & - \frac{\partial \bar{\psi}}{\partial {c}^0_{{}_C}}\,\dot{{c}}^0_{{}_C} + \mathcal{R}_0 \geq 0.
\end{eqnarray}
Introducing the following definitions as in \citet{HOLTHUSEN2023105174},
\begin{eqnarray}\label{mandelbackdefs}
    \bar{\bm{\Sigma}} &:=& 2\,\bar{\bm{C}}_e\,\frac{\partial \bar{\psi}}{\partial \bar{\bm{C}}_e}\nonumber\\
    \bar{\bm{X}} &:=& 2\,\frac{\partial \bar{\psi}}{\partial \bm{C}_g}\,\bm{C}_g\nonumber\\
    \bar{\bm{\Upsilon}} &:=& 2\,\frac{\partial \bar{\psi}}{\partial \bar{\bm{H}}_1}\,\bar{\bm{H}}_1 + 2\,\frac{\partial \bar{\psi}}{\partial \bar{\bm{H}}_2}\,\bar{\bm{H}}_2\nonumber\\
    \bm{G}_1 &:=& \bm{U}_g\,\frac{\partial \bar{\psi}}{\partial \bar{\bm{H}}_1}\,\bm{U}_g\nonumber\\
    \bm{G}_2 &:=& \bm{U}_g\,\frac{\partial \bar{\psi}}{\partial \bar{\bm{H}}_2}\,\bm{U}_g,
\end{eqnarray}
one obtains a compact form of the dissipation inequality in the form as in Eq. \ref{cde_1}. Substitution of the prescribed form of $\bm{U}_g$, i.e.,
\begin{equation}\label{ug_appdx}
    \bm{U}_g := \bm{I} + (\vartheta - 1)\,\bm{\gamma}\otimes\bm{\gamma},
\end{equation}
in the definition of $\bar{\bm{L}}_g$ (Eq. \ref{lgbar}) shall then lead to
\begin{equation}\label{lgbarvg}
    \bar{\bm{L}}_g = \displaystyle{\left(\frac{\dot{\vartheta}}{\vartheta}\right)}\,\bm{\gamma}\otimes\bm{\gamma},
\end{equation}
wherein 
\begin{equation}
    \bm{U}^{-1}_g = \bm{I} - \left(\frac{\vartheta - 1}{\vartheta}\right)\, \bm{\gamma}\otimes\bm{\gamma}
\end{equation}
is utilized. It can additionally be shown that with such a choice for $\bm{U}_g$ as in Eq. \ref{ug_appdx}, 
\begin{equation}
    \bar{\bm{H}}_i = \bm{U}_g\,\bm{H}_i\,\bm{U}_g = \bm{H}_i,\quad i=1,2.
\end{equation}
 With the help of Eqs. \ref{mandelbackdefs} and \ref{lgbarvg}, and the fact that the growth direction $\bm{\gamma}$ is orthogonal to $\bm{a}_{01}$ and $\bm{a}_{02}$ that leads to $\bar{\bm{L}}_g\,\bm{H}_1 = \bar{\bm{L}}_g\,\bm{H}_2 = \bm{0}$, it can be shown that $\bar{\bm{\Upsilon}}:\bar{\bm{L}}_g = 0$. Hence, the dissipation inequality is further reduced from Eq. \ref{cde_1} to the form shown in Eq. \ref{red_cde}.

 \section{Declarations}

\subsection{Funding}
This work has been funded through the financial support of German Research Foundation (DFG) for the following projects:
\begin{enumerate}
    \item ``Drug-eluting coronary stents in stenosed arteries: medical investigation and computational modelling'' (project number 395712048: RE 1057/44-1, RE 1057/44-2).
    \item ``In-stent restenosis in coronary arteries - in silico investigations based on patient-specific data and meta modeling" (project number 465213526: RE 1057/53-1), a subproject of ``SPP 2311:  Robust coupling of continuum-biomechanical in silico models to establish active biological system models for later use in clinical applications - Co-design of modeling, numerics and usability".
    \item ``Modelling of Structure and Fluid-Structure Interaction during Tissue Maturation in Biohybrid Heart Valves'', a subproject of ‘‘PAK961 - Modeling of the structure and fluid–structure interaction of biohybrid heart valves on tissue maturation’’ (project number 403471716: RE 1057/45-1, RE 1057/45-2).
\end{enumerate}
This is gratefully acknowledged.

\subsection{Conflict of interest}
The authors certify that they have no affiliations with or involvement in any organization or entity with any financial interest (such as honoraria; educational grants; participation in speakers’ bureaus; membership, employment, consultancies, stock ownership, or other equity interest; and expert testimony or patent-licensing arrangements), or non-financial interest (such as personal or professional relationships, affiliations, knowledge or beliefs) in the subject matter or materials discussed in this manuscript.

\subsection{Availability of data} 
The data generated through the course of this work is stored redundantly and will be made available on demand.

\subsection{Code availability}
The custom-written routines will be made available on an open platform. The software package FEAP is proprietary and can therefore not be made available.

\subsection{Contributions from the authors}
K. Manjunatha determined the state of the art by reviewing relevant literature, set up governing differential equations, created the user subroutines for implementation in FEAP, interpreted the results, and wrote this article. K. Manjunatha and S. Reese developed the continuum theoretical framework contained within this work. N. Schaaps provided the microscopic and histological data that assisted the mathematical modeling process. M. Behr and F. Vogt participated in project discussions, provided conceptual advice, gave feedback on the interpretation of results, read the article, and provided valuable suggestions for improvement. All the authors approve the publication of this manuscript. 

\bibliography{isr_pp_paper}
%\printbibliography

\end{document}